# Order-disorder duality of high entropy alloys extends non-linear optics

Valentin A. Milichko,[1,†] Ekaterina Gunina,[2] Nikita Kulachenkov,[3] Maxime Vergès,[1] Luis Casillas Trujillo,[4] Maria Timofeeva,[5] Jaafar Ghanbaja,[1] Stéphanie Bruyère,[1] Andrey Krasilin,[2] Mikhail Petrov,[2] Michael Feuerbacher,[6] Yann Battie,[7] Rachel Grange,[5] Jean-Pascal Borra,[8] Jean-François Pierson,[1] Vincent Fournée,[1] Thierry Belmonte,[1] Janez Zavašnik,[9,10] Björn Alling,[4] Joseph Kioseoglou,[11,12] Ivan Iorsh,[13] Julian Ledieu,[1] Uroš Cvelbar,[9,†] Alexandre Nominé[1,9,14,†]

[1] Université de Lorraine, Institut Jean Lamour, CNRS, 54000 Nancy, France

[2] ITMO University, Saint Petersburg, Russia

[3] The Rockefeller University, New York, NY 10065, USA

[4] Department of Physics, Chemistry and Biology (IFM), Linköping University, Linköping 58183, Sweden

[5] ETH Zurich, Department of Physics, Institute for Quantum Electronics, Optical Nanomaterial Group, 8093 Zurich, Switzerland

[6] Ernst Ruska-Centre for Microscopy and Spectroscopy with Electrons, Forschungszentrum Jülich GmbH, 52425 Jülich, Germany

[7] Laboratoire de Chimie et Physique, Approche Multi-Echelle des Milieux Complexes, Univ. Lorraine, Metz, France

[8] Laboratoire de Physique des Gaz et des Plasmas, UMR8578 CNRS, Université Paris Sud, Université Paris Saclay, France

[9] Department of Gaseous Electronics, Jožef Stefan Institute, Jamova cesta 39, SI-1000 Ljubljana, Slovenia

[10] Max-Planck-Institut für Eisenforschung GmbH, Max-Planck-Straße 1, 40237 Düsseldorf, Germany

[11] Department of Physics, Aristotle University of Thessaloniki, GR-54124 Thessaloniki, Greece.

[12] Center for Interdisciplinary Research and Innovation, Aristotle University of Thessaloniki, Thessaloniki, Greece.

[13] Department of Physics, Engineering Physics and Astronomy, Queen's University, Kingston, Ontario K7L 3N6, Canada

[14] LORIA, Université de Lorraine, CNRS, INRIA, Nancy, France

[†] Corresponding authors : valentin.milichko@univ-lorraine.fr, alexandre.nomine@univ-lorraine.fr, uros.cvelbar@ijs.si

**Order *vs* disorder in the structure of materials plays a key role in the theoretical prediction of the properties. However, this structural description appears to be ineffective for new families of materials such as high entropy alloys (HEAs) which combine crystallographic order with chemical disorder. Here, we demonstrate for five-element HEAs as pure solid-solutions that the chemical disorder of the elements, decorating their cubic structure, underlies the generation of second optical harmonics, which overcome the theoretical limit imposed on centrosymmetric crystals. Moreover, we discover that this disorder, inherent to HEAs, sets a threshold non-linear light emission of $4^{th}$ to $26^{th}$ order. As a consequence of the 0.5 eV broadening of the energy levels of five elements of the HEA, the emission spectrum covers broad visible (400 – 650 nm) and infrared (800 – 1600 nm) ranges. In addition to the challenge of theoretically predicting non-**

**linear effects in unconventional materials, the duality of 'structural order and chemical disorder' in HEAs offers the opportunity to design sustainable alternatives to urgently needed optical materials.**

Nonlinear optical materials have a variety of applications in modern technology to manipulate and control light.[1–4] These materials generate new frequencies of light, convert one wavelength to another, or modulate the intensity or phase of light.[1,5] In generally, crystal structure and symmetry form the basis for certain non-linear optical effects. For example, the absence/presence of inversion symmetry in a crystal (also known as centrosymmetry or $\bar{x}, \bar{y}, \bar{z}$ symmetry) is used to predict the existence of second/third order optical nonlinearities, respectively.[6]

The centrosymmetry of a crystal is usually easy to assess in conventional materials with a defined structure, and space group (Figure 1). However, for solid-solution alloys such as high entropy alloys (HEAs), where a periodic lattice is randomly occupied by different atoms, this is controversioal (Figure 1). In other words, HEAs combine crystallographic order and chemical disorder.[7,8] Consequently, the definition of a unit cell, which could be extrapolated to the whole material by translational symmetry, is no longer possible. Intriguingly, HEAs exhibit long-range symmetry as evidenced by discrete X-ray diffraction patterns; however, short-range symmetry is impossible to define due to the randomness of the population of crystallographic sites. This duality rises a challenge for the theoretical framework of various non-linear (optical, electronic, mechanical, and others) effects.

This duality highlights the difficulty of categorizing HEA as either centrosymmetric or non-centrosymmetric material, making it therefore challenging to predict *a priori* second or third order nonlinearities.[6] Third-order nonlinear effects have also been reported in HEAs—for example, FeCoNiAlTi powders exhibiting strong $\chi^3$ responses when used as broadband saturable absorbers in ultrafast fiber lasers[9]. Notably, third-harmonic generation and related $\chi^3$ processes do not require broken inversion symmetry and can occur in centrosymmetric crystals. While diffraction confirms an average fcc lattice, local chemical environments are far from unique. Cantor recently estimated that a five-component fcc HEA can host on the order of $10^{13}$–$10^{14}$ distinct nearest-neighbour clusters[10]—an enormous configuration space even if some degree of chemical short-range order (CSRO) is present. Such a landscape means that many nanoscale regions inevitably differ in symmetry from the average lattice. Whether these sub-wavelength variations influence nonlinear optical responses is precisely the question addressed in this work.

To investigate this peculiarity experimentally, a second-order non-linear optical effect (second harmonic generation, SHG) was studied for a representative FeCoCrMnNi HEA *aka* Cantor alloy (with equiatomic face-centred cubic, fcc, centrosymmetric structure)(Figure 2a). Figure 2b shows the presence of SHG signal centred at 525 nm from the FeCoCrMnNi target, pumped by 150 fs light pulses with a wavelength 1050 nm (Supplementary Figure S3). It should be noted that the contribution of the grain boundaries of the FeCoCrMnNi target to its SHG signal is minimised, as the crystallite size is millimetric,[11] *i.e.* the concentration of grain boundaries is very low. Taking into account poor plasmonic properties of the surface of the FeCoCrMnNi target (Supplementary Figure S2), the SHG signal should mainly originate from the lattice distortion in the crystal. Although, the atoms of the Cantor alloy have similar atomic radii (from 124 to 128 pm), the density functional theory (DFT) calculations (Figure 2e,f, and Supplementary Note 6) showed a distortion at d = 3.5 pm (Figure 2e), which is consistent with the values measured experimentally by neutron diffraction.[12] To enhance the lattice distortion, we used a body-centred cubic (bcc) $AlCrFeNiTi_{0.21}$ alloy consisting of three smaller atoms (Ni, Fe, and Cr with metal radii of 124, 126, and 128 pm, respectively), and two larger atoms (Ti and Al, with corresponding metal radii of 147 and 143 pm, see Figure 2d). Consequently, DFT calculations (Figure 2e,f, Supplementary Note 6) confirmed that significant differences in atomic radii between the smaller and larger atoms of $AlCrFeNiTi_{0.21}$ induce substantial lattice distortions

(maximum probability at d = 13 pm). A comparative analysis with the Cantor alloy (Figure 2c) under the same conditions revealed that AlCrFeNiTi$_{0.21}$ generates five times more intense SHG signal. The quadratic slope of the SHG intensity (Figure 2c, inset) and the exact position of its wavelength also confirm the second-order nonlinear process. It is worth noting that the calculated lattice distortion for both HEA targets can be approximately compared to the displacement of atoms in a unit cell of conventional non-centrosymmetric crystals such as $KH_2PO_4$ (KDP),[13] for which d equals 1 to 5 pm, being a key factor for the observation of second-order nonlinear effects.[14]

The increase in the atomic radius difference in the alloy AlCrFeNiTi$_{0.21}$ promotes the compositional heterogeneity of the alloy[15] with two regions, one richer in Cr and Fe with a lattice constant of 289.9 pm, and the other richer in Al, Ni, and Ti with a lattice constant of 291 pm. This heterogeneity is related to the eutectic structure and spinodal decomposition, highlighting that the microstructure of the alloy can also contribute to the SHG. Thus, as in mechanics, a "cocktail effect"[16] modulates the efficiency of the initial SHG in HEA, similar to defect engineering of semiconductors for nonlinear optics.[39]

In addition to the observed coherent SHG process, we discovered that pumping HEA targets with 525 nm light (see Figure 3a and Methods) leads to incoherent broadband light emission. Figure 3b shows the emission spectrum of the Cantor alloy ranging from 800 nm to at least 1650 nm (i.e., 0.8 eV broadening), limited by the sensitivity of the detector (Supplementary Figure S5). This spectral range is of particular interest as it lies within both the biological (900 to 1625 nm) and the telecommunication (1260 to 1625 nm) windows. Moreover, the spectrum fund covers most of the emission spectra of rare-earth elements in the infrared range (Supplementary Figure S10). A similar emission spectrum is also detected for the AlCrFeNiTi$_{0.21}$ target (Supplementary Figure S6). Here, the shift of laser pumping wavelength from 525 to 350 nm resulted in light emission from 420 to 650 nm (Supplementary Figure S7, for AlCrFeNiTi$_{0.21}$ and Cantor targets), again corresponding to a spectral width of 1 eV. The estimated efficiency of such broadband light emission ($10^{-4}$ to $10^{-3}$, Figure 3c) is comparable to that of conventional gold nanoclusters, but significantly lower than that of organic materials or complex semiconductor structures (see also Supplementary Figures S11, S12). The latter can be explained by the inverse relationship between the emission efficiency and the width of the spectrum,[17] which makes it challenging to realize simultaneously such broadband emission with high efficiency for a specific material.

A more detailed analysis of the incoherent broadband light emission reveals that the dependence of the emission intensity $I$ and the shape of the emission spectrum on the pump fluence $F$ is strongly non-linear. Initially, there is no light emission at a low value of $F$ (Figure 3d). An increase in $F$ leads to a threshold value for the emission with a subsequent increase in its intensity. We have found that $I$ is $\sim F^n$, with the power $n$ varying between 4 and 26 (see below, Supplementary Figures S6b, S19). Thus, the light emission could be a reversible process (Figure 3e) with no any changes of the surface of HEA targets after exposure to laser pump. Secondly, a further increase in $F$ results in both a distortion of the shape of the emission spectrum (Supplementary Figure S18b) and irreversible changes on the surface of the HEA (Supplementary Figure S9): At a constant high value of $F$, the emission intensity varies over time and eventually disappears (Figure 3d).

To elucidate the nature of incoherent broadband light emission from HEA, a blackbody emission effect (BBE) should be excluded. According to Wien's law, the light emission at 1400 nm (Figure 3b) would lead to an HEA temperature of 2070 K, which is higher than the melting point of both of our alloys (see Supplementary Figure S13). Furthermore, the BBE hypothesis combined with the constant emission peak when varying the value of $F$ would lead to the unreasonable conclusion that the temperature increase of HEA is independent of the pump laser fluence. Furthermore, it would be conceivable that the heating of the HEAs by laser pumping oxidizes them; so that the broadband light emission would originate from an oxide state of the metals on the target surface. However, this hypothesis seems unlikely due to the irreversibility of the oxidation process, which contradicts the observed reversible light emission in Figure 3e. However, oxidation occurs at extremely high $F$ values, which affect the shape of the emission spectrum (Supplementary Figure S18b) and the color of the target surface (Supplementary Figure S9).

The preliminary conclusion incoherent broadband light emission is still inherent to HEAs, leads us to the problem of describing the physical mechanism of photoluminescence (PL) of metals. Although such PL has been known for decades, its underlying nature remains unidentified, with two potential mechanisms being considered, such as hot electron inter-intra band transitions[18–20] and resonant Raman scattering.[21,22] In our case, the conversion of 2.37 eV (i.e., 525 nm wavelength) incident photons into 0.75 - 1.55 eV re-emitted photons is consistent with metal PL. Also, the observed strongly nonlinear ($I \sim F^n$) behavior of the light emission intensity for HEA targets can possibly be attributed to the process of multiphoton absorption.[23] However, absorption of at least four photons (each with an energy of 2.37 eV) would largely overcome the ionization energy level of all five elements of HEA (6.77 to 7.9 eV).[24] In contrast, such slopes are typical of photon avalanching,[25] which has recently been observed in semiconducting nanoparticles.

The electronic band-structure of the Cantor alloy was thus determined with the air of first-principles calculations (see Supplementary Note 5). Figure 4 demonstrates that the structure has both strongly broadened (by app. 0.5 eV) and sharp bands. For example, near to the L-point in the (1,1,1)-direction in the fcc lattice, we see both a sharp energy band at the Fermi level and a strongly broadened energy band about 1 eV above the Fermi level. On the other hand, at the Γ-point, all bands near the Fermi Energy $E_F$, are quite diffuse and broad. At the U, K point, we again see a rather sharp band below $E_F$, while the bands above $E_F$ are very broad. We hypothesize that the reason for the broadening of the bands is the chemical disorder and difference of electronegativity, augmented by the spin-disorder in the disordered paramagnetic state. Strictly speaking, the translational symmetry of HEA is broken and

*k* is no longer a good quantum number. In practice, this means that despite the differences between five chemical elements, the preservation of bands due to their open 3d-shells results in varying degrees of broadening, which indicates finite and non-constant lifetimes. More intriguing is that the band at $E_F$ at the L-point still remains practically intact and sharp. Regardless of the origin of the latter, the combined consequence is that both the dark state, and broadened levels are present in the system which together may be at the origin of the experimentally observed effect of both presence of 0.8 – 1 eV broadband light emission, as well as the broadening of the core level spectra of the Cantor alloy.

To demonstrate that the emission of coherent and incoherent light by HEAs is independent of their size, we fabricated microscale particles of Cantor alloy by electric discharges in dielectric liquid (EDDL, see Methods and Supplementary Note 3). The resulting 0.1 to 2 µm HEA microparticles (Figure 5a,b) show an SHG signal, which was confirmed by two independent methods (Fig. 5d and Supplementary Figure S16). A single Cantor alloy microparticle demonstrates a conventional dipole-shape polarisation-dependant SHG signal (Figure 5c) with a 40-50° angular broadening due to the lattice distortion. Since the Cantor alloy is a metal that has the potential to exhibit plasmon resonance (albeit blurred by large optical losses, see Supplementary Figure S2), the surface effect of the microparticles could contribute to the detected SHG signal. Therefore, we analyzed the dependence of SHG on the microparticle size (Supplementary Figure S17) and found no direct quadratic or cubic (for the volume contribution) dependence. On the one hand, the latter may be a consequence of the SHG due to the microparticle volumes because of the significant penetration depth of the 1050 nm pumping light; on the other hand, it does not completely exclude the contribution of the surface. The incoherent broadband light emission with the threshold was also demonstrated for Cantor microparticles. Figure 5e shows the emission spectra whose shapes match well with those of the target (Figure 3b, and Supplementary Figures S8, and S18). In contrast to the target, however, the microparticles exhibit a steeper increase in light emission intensity ($I \sim F^n$): 8 to 26 values of *n* can be observed depending on the size of the microparticles (Figure 5f), while no emission was observed for the microparticles with a size of less than 1 μm (Figure 5g).

Optical and numerical experiments demonstrate that two intrinsic properties of HEAs (i.e., the differences in atomic radius and electronegativity) determine the occurrence of non-linear optical responses. In particular, the differences in atomic radius lead to lattice distortion, which is responsible for SHG, while the differences in electronegativity broaden the electronic levels and contribute to broadband light emission. Since the emission behavior is related to the electron band structure, it is anticipated that altering the alloy composition could enable modulation of broadband emission spectra, offering the potential for tailored emitters with tunable spectral range (from the visible to the infrared). The compositional diversity of HEAs is enormous: for example, 30,201 equimolar quinary HEAs have been numerically predicted from a pool of 40 elements,[26] out of a staggering 658,000 possible combinations. Taking into account the possibility of synthesizing non-equimolar alloys, this compositional space increases exponentially. With 40 elements, the number of potential alloys increases from 658.000 equimolar combinations to $10^{36}$ for a 1 % compositional step.[27] Even if only a fraction of these alloys form single-phase HEAs (4.6 % were reported by Chen *et al*[26]), the number of possible candidates remains astronomical, surpassing the total number of stars in the universe. The discovery thus sets the stage for theoretical studies of systems that combine chemical disorder with crystallographic order (such as solid solutions).

Apart from the obvious scientific challenge, the wised range of potential materials for infrared emissions must also be seen as an opportunity to improve both the performance and sustainability of these materials. It is worth noting that HEAs outperform the known infra-red emission materials in terms of emission bandwidth (Si, InGaAs, and rare-earth based materials, see Figure 6 and Supplementary Note 7). Furthermore, the carbon footprint associated with the metal extraction of FeCoCrMnNi and AlCrFeNiTi$_{0.21}$ is one and two orders of magnitude lower than that of InGaAs and rare-earth based infrared emitters, respectively (Figure 6a). Thanks to the abundance of the constituting materials, mass-scale production is possible, while remitters based on rare-earths are limited to a few dozen to hundreds of tons per year (Figure 6b). The two alloys studied consist largely of primary mined metals, which is not the case for InGaAs and rare-earth elements for which the companionability is very high[28] (Figure 6c). Finally, the supply risks of the investigated HEAs are significantly lower than for conventional infra-red emitters (Figure 6d). Thus, we demonstrate again that HEA is not only an unconventional material for extension of fundamental description of non-linear effects, but also opens up a new avenue for sustainable materials for non-linear optics.

**Methods**

**Materials.** AlCrFeTiNi target ($Al_{0.2375}Cr_{0.2375}Fe_{0.2375}Ni_{0.2375}Ti_{0.05}$) has bee purchased from Plansee GmbH (Material No.: 12578388; Batch No.: Al7526; packaging date: 09.03.2021). The Cantor alloy was synthesized from pure powders by plasma sintering.

**Optical measurements.** SHG and broadband light emission analysis were performed under ambient air conditions using a home-made setup that allows optical measurements to be performed in reflection and scattering geometry for HEA targets and single microparticles (Supplementary Figures S3, S4). As a pumping light source, femtosecond laser system (TEMA, Avesta Project), emitting 150 fs laser pulses at a wavelength of 1050 (for SHG) and 525, 350 nm (to initiate the broadband light emission) with a repetition rate of 80 MHz was utilized. The optical path of the setup also included a set of mirrors (Thorlabs protected silver mirrors with a reflectance of 97.5% in the (0.45;2) µm region), filters (Thorlabs FELH0600, FESH 0600 allowing to cut the pumping wavelength), and dividing cubes. Mitutoyo NUV 20x (NA=0.42), and Mitutoyo NIR 50x (NA=0.42) objectives were used to initiate SHG and broadband light emission; while Mitutoyo M Plan Apo NIR 50x (NA=0.42) objective was used to collect the optical signal. The spectrometer used was the commercially available Horiba LabRam confocal spectrometer with thermoelectrically cooled charge-couple device (Andor iDus InGaAs) and 600, 150 g mm$^{-1}$ diffraction grating, has been used. For precise positioning of individual Cantor microparticles for SHG and light emission analysis, a Thorlabs piezo stage with 10 nm step over x,y,z axes and 50x50 µm range was also used.

**Light emission efficiency.** Evaluation of the broadband light emission efficiency (Supplementary Figures S11, S12, and S19) was carried out by dividing the intensity of the detected optical signal by that of the pumping signal, taking into account the following factors: Detector (Andor iDus InGaAs) sensitivity in the range of 523.5 and 800-1650 nm; the transmission coefficients of the filters (FELH0600, FESH0600) used; transparency windows of the objectives and reflectivity values of silver mirrors at different wavelengths; signal accumulation time (0.1 second for pumping, and 10 seconds for HEA targets and the Cantor microparticles); absorption coefficients of the Cantor at a wavelength of 523.5 nm; an angle of collection of the emitted light by the objective; and also non-absorbed/reflected pumping energy due to weak focusing of pumping light by a low-aperture objectives.

**Nanoparticle fabrication (Electric Discharges in Dielectric Liquid).** The experimental setup (Supplementary Figures S14) consists in immersing of Cantor alloy electrodes in liquid nitrogen and applying 10 kV and 500 ns pulses during 600 ns. DC voltage power supply (Technix SR15-R-1200–15 kV–80 mA) was connected to a solid-state switch (Behlke HTS-301-03-GSM) connected to one pin-electrode, while the other pin-electrode was grounded. The voltage rise time was 20 ns. The pulse frequency was set to 10 Hz during 30 minutes. The inter-electrode gap distance remained constant at 100 µm. Further details on the synthesis can be found in Hamdan *et al.*[29]

**Theory.** The DFT calculations were performed using the projector augmented wave method as implemented in the Vienna Ab-initio Simulation Package (VASP)[30,31] with the Perdew-Burke-Ernzenhof generalized gradient[32] to model the exchange correlation effects. All calculations were performed using a 500 eV kinetic-energy cut-off, and with a convergence criterion on forces of 0.01 eV Å$^{-1}$ for structural relaxations. The multi-component alloy was modelled using fcc underlying crystal lattice, equimolar concentrations, and a supercell with 80 atoms, corresponding to 16 atoms of each element. The supercell was created using the special quasirandom structure method (SQS).[33] The supercell was first optimized with respect to volume using fixed atomic position. The equilibrium structure was obtained by relaxing the supercell with a ferromagnetic state, allowing atomic displacements and supercell volume change. The CLS and bands were calculated with a paramagnetic state simulated with a collinear supercell implementation[34] of the disordered local moment (DLM) approximation[35] by superimposing the DLM on the ferromagnetically relaxed structure and no further relaxations were performed.

The core level shifts were simulated using the Z+1 approximation.[36] The core ionized atom Z* is replaced by the next element in the periodic table: Z+1. In this way, the core-electron excitations and the effect of a core-level hole screened by an extra valence electron are approximated with an extra proton and an extra electron. The photoelectron binding energy shifts, relative elemental references, can be calculated for all atoms in the modelled multi-component alloy supercell.[37] The band unfolding to obtain the density of states was performed using the BandUp code.[38,39]

The VASP ab-initio simulation package[31,31] is be implemented in order to reveal favorable structures of $Al_{0.2375}Cr_{0.2375}Fe_{0.2375}Ni_{0.2375}Ti_{0.05}$ and $Fe_{0.2}Co_{0.2}Cr_{0.2}Mn_{0.2}Ni_{0.2}$ HEAs. The core argument of HEAs is that the configurational entropy is expected to be exceptionally high. As a result, when these metallic structures form at elevated temperatures, they should exhibit a significant degree of randomness.

Calculations are performed under the GGA-PBE approximation, and the standard PAW VASP 6.4 pseudopotentials were used. Except for the Al and Co atoms where the standard pseudopotentials are used, for all the rest atoms the "_pv" pseudopotentials are used, which implies that the p semi-core states are treated as valence states.

The HEAs are modeled, initially for 3x3x3 unit cells leading in supercells of 216 atoms for the bcc structure of the $Al_{0.2375}Cr_{0.2375}Fe_{0.2375}Ni_{0.2375}Ti_{0.05}$ and in supercells of 432 atoms for the fcc structure of $Fe_{0.2}Co_{0.2}Cr_{0.2}Mn_{0.2}Ni_{0.2}$. The initial structural models are formulated with lattice constants linear combination of the atomic compositions, considering the individual elements relaxed metallic structures. Afterwards a 3x3x3 Γ-centered k-point mesh is be used. Also, metaGGA functionals in the form of r2SCAN calculations are considered. The r2SCAN structures are obtained from a two-step workflow which is comprised of the initial GGA-PBE optimization, followed by final optimization with r2SCAN. In all the cases spin-polarized calculations are used.

The process required two different steps. The initial relaxation of the atomistic models is performed with ISIF=3. This setting enables the calculation of both forces and the stress tensor. Also, all the possible degrees of freedom that are allowed to change during the relaxation process (ionic positions, cell volume, cell shape) are activated and used. The second relaxation of the atomistic models is conducted with ISIF=1. This relaxation considers the forces and only the trace of the stress tensor, with the positions of the ions being the only degrees of freedom that change. The plane wave basis energy cutoff is 520 eV to compensate for volume relaxations. The smearing method used for GGA-PBE calculations is the Methfessel-Paxton method. Finally, OVITO basic 3.8.5[40] was used to visualize the atomistic structures.

**Acknowledgements**

VM acknowledges the Chaire de Professeur Junior (CPJ) project from Ministre de l'enseignement supérieur et de la recherché, France. AN, JZ and UC are grateful to Slovenian Research and Innovation Agency (ARIS) program P1-0417 and project J2-4440 for financial support. AN and JFP are grateful to Institut Carnot ICEEL. VF and JL contributions was partially supported by the European Integrated Center for the Development of Metallic Alloys & Compounds (ECMetAC). JK acknowledges the computational time granted from the Greek Research & Technology Network (GRNET) in the "ARIS" National HPC infrastructure under the project NOUS (pr015006). MF is grateful to the German Research Foundation (DFG) for financial support through the Priority Program SPP 2006 "Compositionally Complex Alloys – High Entropy Alloys (CCA-HEA)" under grant no. FE 571/4-2. BA acknowledges financial support from the Swedish Research Council (VR) through Grant No. 2019-05403 and 2023-05194, from the Swedish Government Strategic Research Area in Materials Science on Functional Materials at Linköping University (Faculty Grant SFOMatLiU No. 2009-00971). The computations were enabled by resources provided by the National Academic Infrastructure for Supercomputing in Sweden (NAISS) partially funded by the Swedish Research Council through grant agreement no. 2022-06725. The authors are very grateful to Prof. I.A. Abrikosov for fruitful discussions.

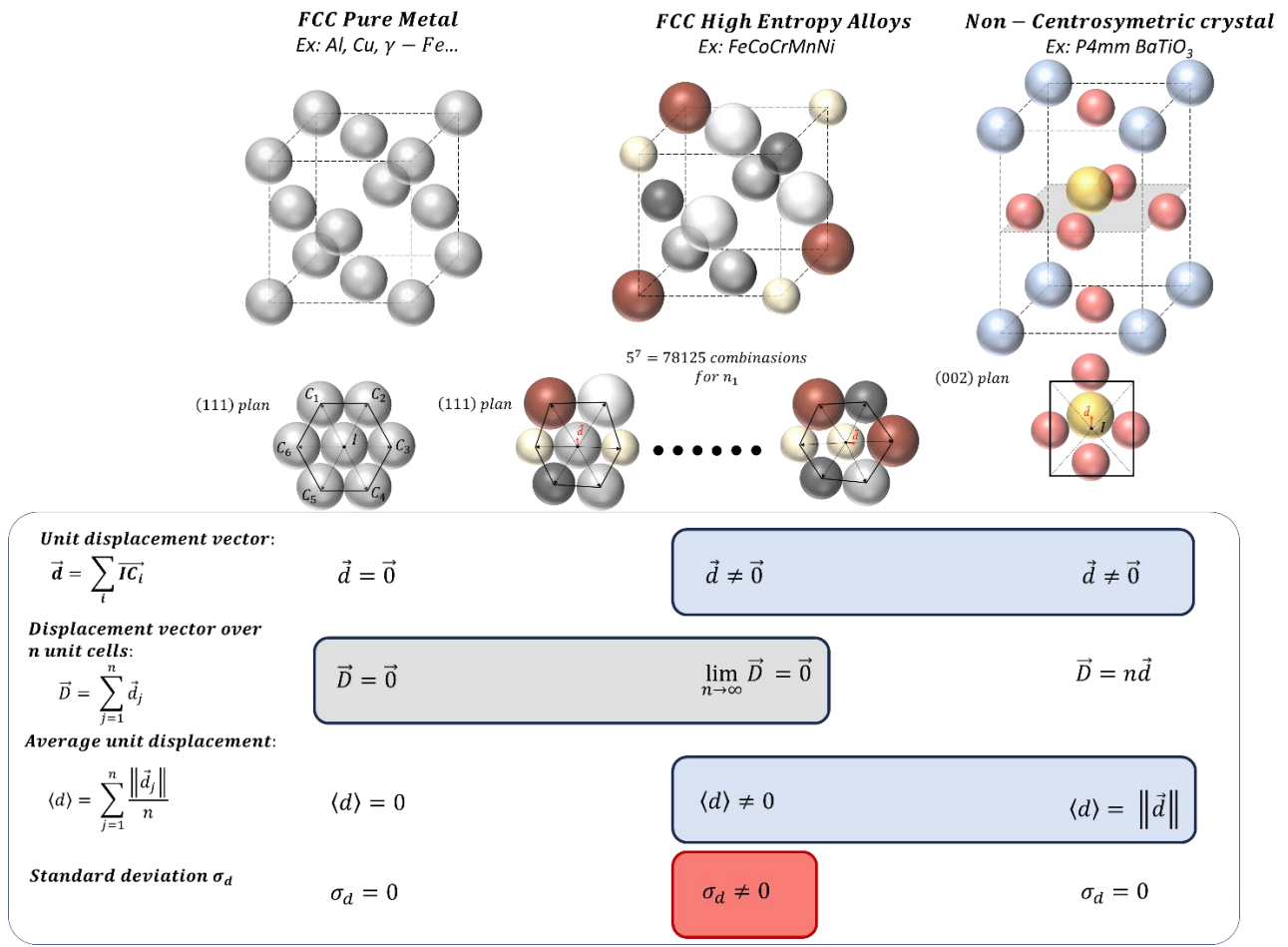

**Figure 1. Concept of order-disorder duality of HEA.** Structure representation of a typical single-element face-centered cubic (fcc, i.e. centrosymmetric) metal, fcc HEA, and a non-centrosymmetric crystal (*P4mm*) Baryum titanate. The displacement vector $\vec{d}$ can be expressed as the sum of the vectors $\vec{IC}$ between the center of inversion I of the average lattice (e.g. fcc lattice with a = 0.365 nm for FeCoCrMnNi and measured by diffraction), and the atom centers with coordinate 1 ($n_1$). The displacement vector reflects the non-centrosymmetricity and shows that fcc HEA shares some features of both the non-centrosymmetric materials (non-null unit displacement and average displacement) and the centrosymmetric materials (null displacement over a large number of unit cells) and some unique feature such as non-null standard deviation of the displacement vector over n unit cells.

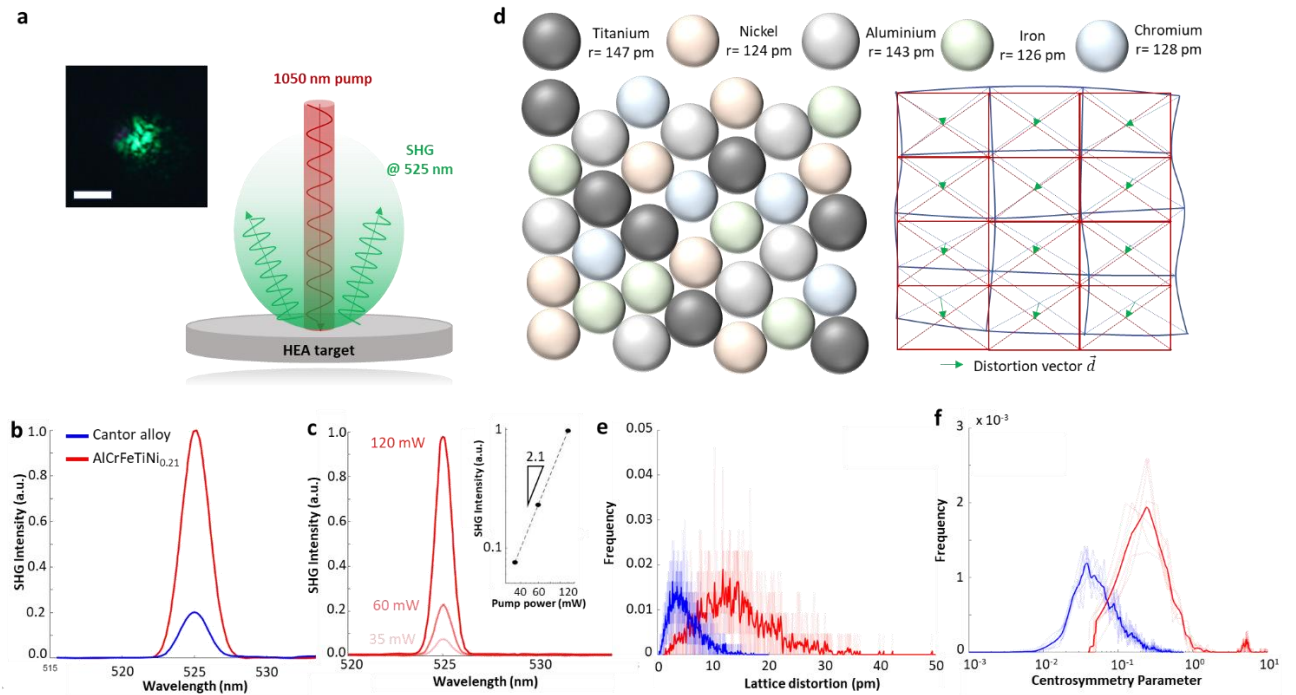

**Figure 2. SHG of HEA. a,** Schematic of the SHG experiment with HEA target (AlCrFeNiTi$_{0.21}$ or Cantor alloy). Inset: Optical image of the SHG signal from the surface of the Cantor alloy pumped with 1050 nm light (scale bar 20 µm). **b,** Comparison of SHG intensity originating from the Cantor and AlCrFeNiTi$_{0.21}$ targets measured under the same conditions. **c,** Normalized SHG spectra from AlCrFeNiTi$_{0.21}$ target with corresponding quadratic slope over pumping power. **d,** Schematic of the lattice distortion of AlCrFeNiTi$_{0.21}$ due to the difference in atomic radii of Al and Ti on a one hand and Fe, Cr and Ni on the other hand. **e,f** Lattice distortion and centrosymmetric parameter calculated for 3x3x3 unit cells by DFT.

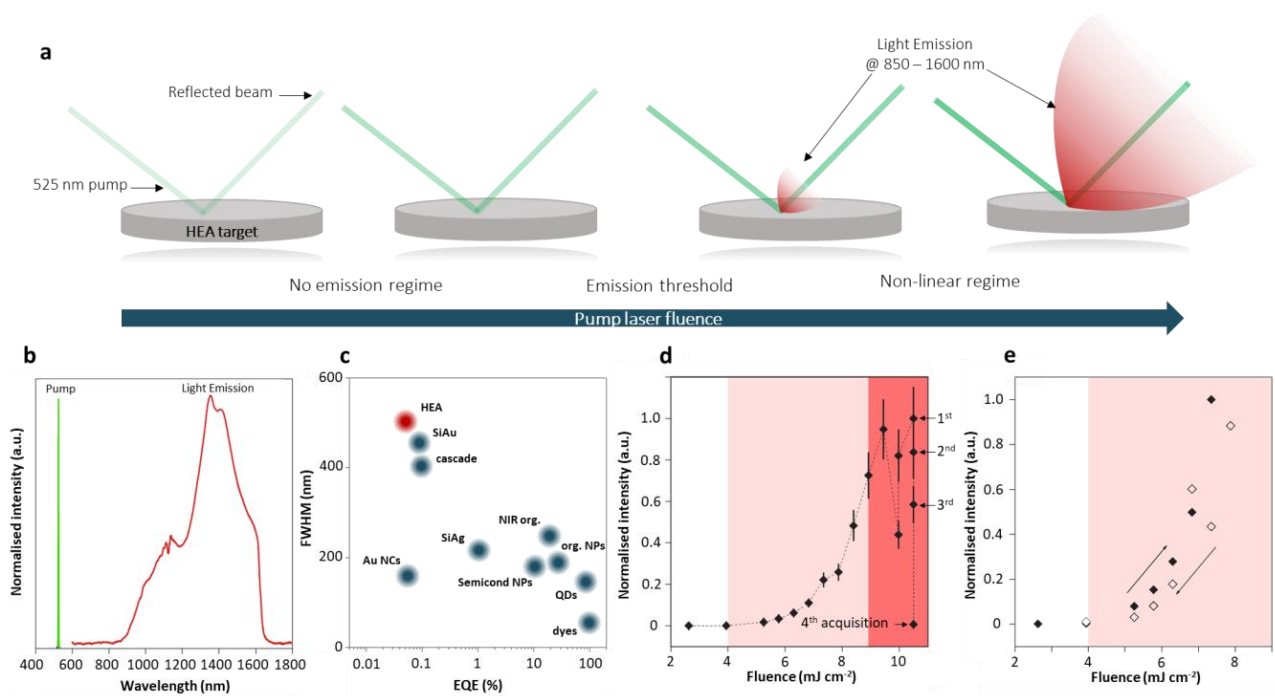

**Figure 3. Broadband light emission from HEA. a,** Schematic of optical measurement of incoherent broadband light emission from the HEA targets pumped by 525 nm light, showing different emission regimes with pumping fluence. **b,** Normalized spectra of pump and light emission from the Cantor alloy target. **c,** Estimated quantum efficiency (EQE) and bandwidth (FWHM) of the broadband light emission for HEA and other materials (see Supplementary Figure S12). **d,** Dependence of the broadband emission intensity on pumping fluence, displaying three different regimes: no-emission at 0-4 mJ cm$^{-2}$, emission threshold at 4 mJ cm$^{-2}$ followed by non-linear increase in emission intensity from 4 to 9 mJ cm$^{-2}$, and the laser damaging regime at 9 mJ cm$^{-2}$. **e,** Reversibility of the broadband light emission over the limited pumping fluence range.

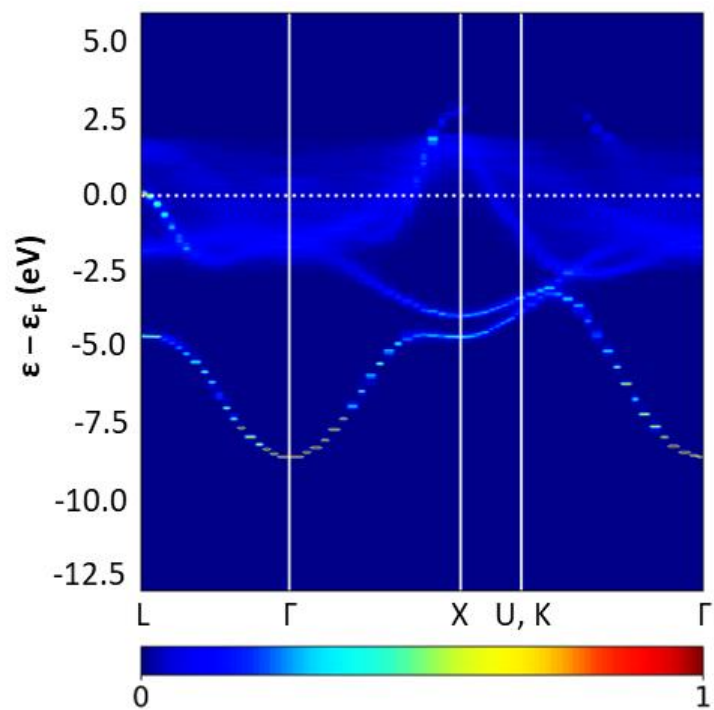

**Figure 4. Electronic band structure of the Cantor alloy.** The structure displays the simultaneous presence of both highly broadened and sharp energy bands. Electronic band structure was calculated using DFT for 80 atoms (see Supplementary Note 5, and Methods).

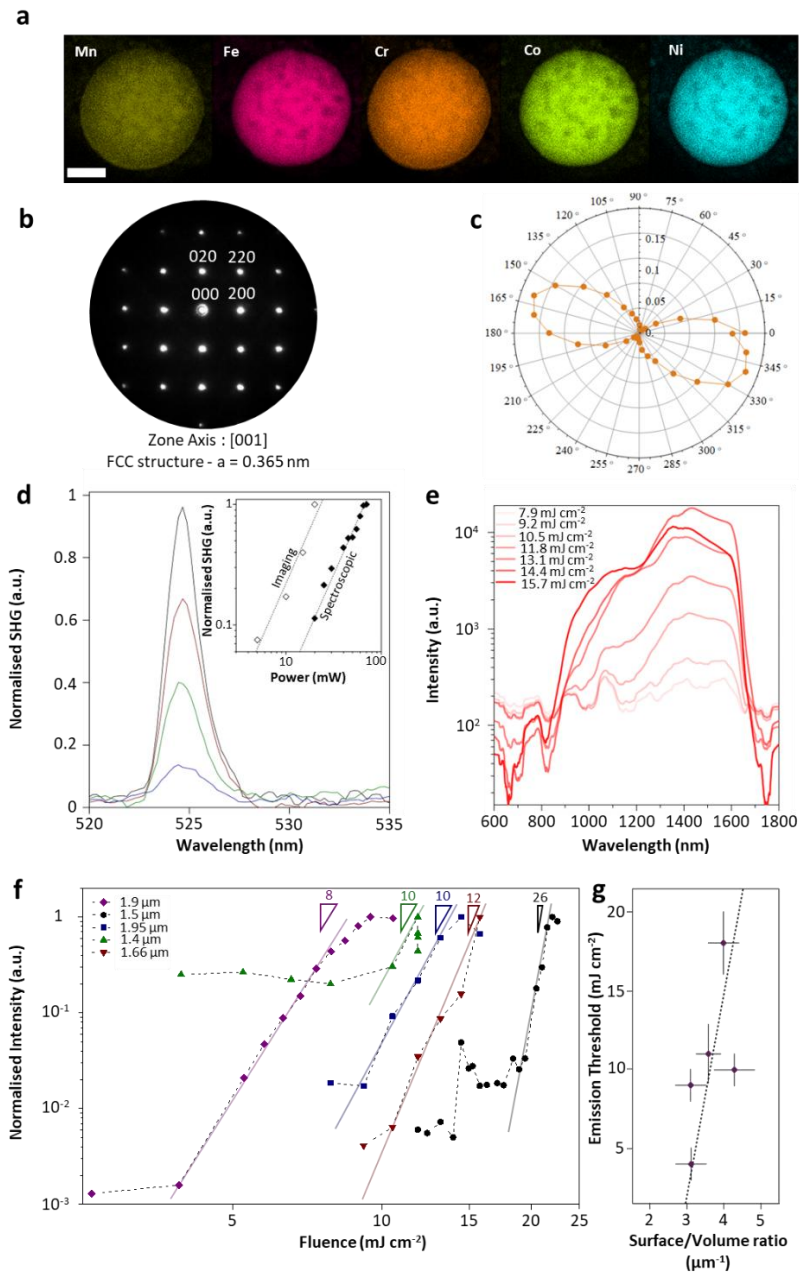

**Figure 5. Coherent and incoherent light emission by Cantor microparticles. a,** Elemental mapping of a Cantor alloy microparticle synthesized by electric discharges in dielectric liquid. Scale bar, 50 nm. **b,** Electron diffraction pattern in [001] zone axis confirming fcc structure with a lattice parameter of 0.365 nm. **c,** Polarization dependent SHG signal from a single Cantor alloy microparticle. **d,** Normalized SHG signal from a single Cantor alloy microparticle with corresponding quadratic slope confirmed by two independent methods (see Methods, and Supplementary Figure S16). **e,** Broadband light emission spectra from a single Cantor alloy microparticle pumped by 1050 nm light at different fluence. **f,** Normalized intensity of broadband light emission from a single Cantor microparticle of different diameter revealing 8 – 26$^{th}$ order of light emission. **g,** Dependence of the threshold pumping fluence for the light emission on the aspect ratio of corresponding Cantor alloy microparticles.

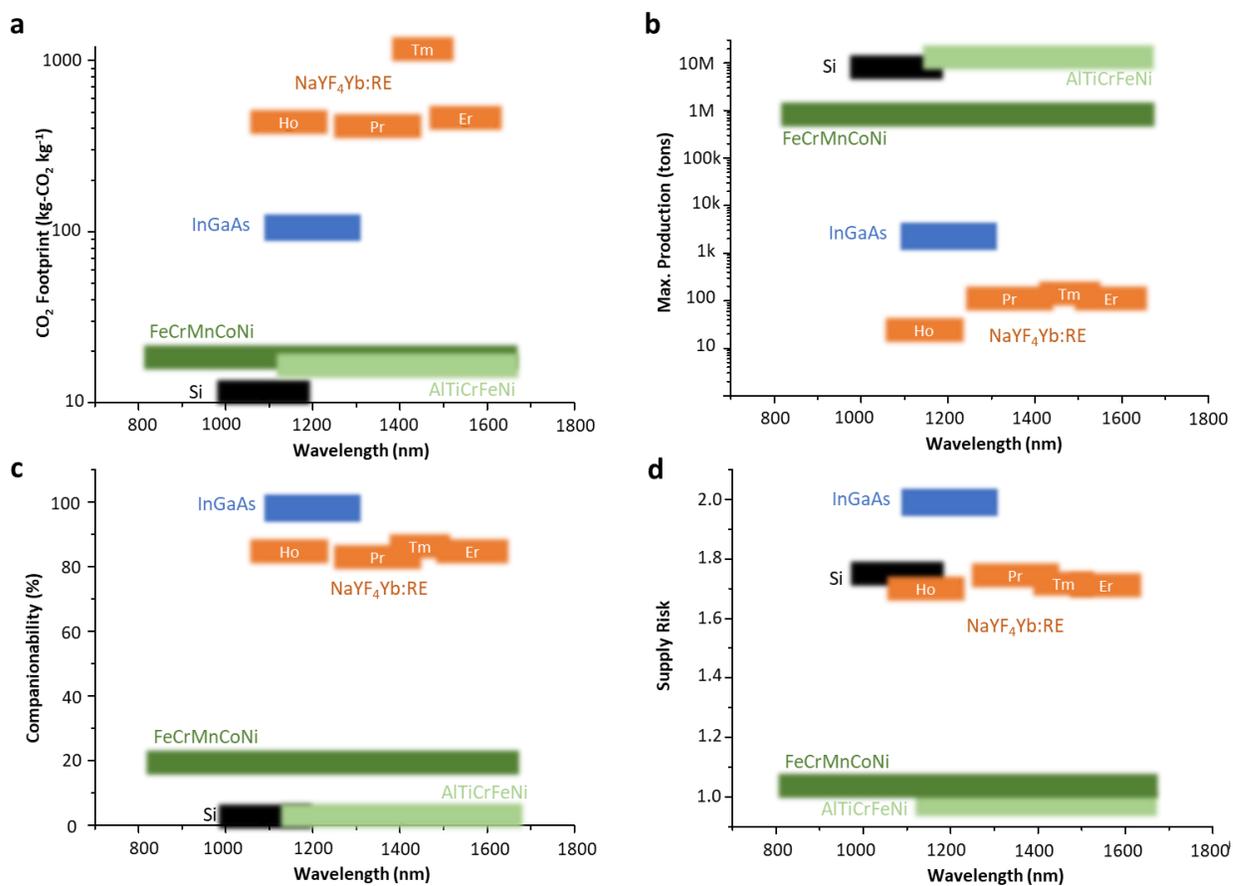

**Figure 6. Sustainability *vs* emission bandwidth of HEA and conventional infrared light emitting materials, showing our material positioning (green)**. **a,** Primary production $CO_2$ footprint, **b,** maximum production, **c,** average companionability level, and **d,** supply risk of different light emitters (see Supplementary note 7).

**Order-disorder duality of high entropy alloys extends non-linear optics**

Valentin A. Milichko,[1,†] Ekaterina Gunina,[2] Nikita Kulachenkov,[3] Maxime Vergès,[1] Luis Casillas Trujillo,[4] Maria Timofeeva,[5] Jaafar Ghanbaja,[1] Stéphanie Bruyère,[1] Andrey Krasilin,[2] Mikhail Petrov,[2] Michael Feuerbacher,[6] Yann Battie,[7] Rachel Grange,[5] Jean-Pascal Borra,[8] Jean-François Pierson,[1] Vincent Fournée,[1] Thierry Belmonte,[1] Janez Zavašnik,[9,10] Björn Alling,[4] Joseph Kioseoglou,[11,12] Ivan Iorsh,[13] Julian Ledieu,[1] Uroš Cvelbar,[9,†] Alexandre Nominé[1,9,14,†]

[1] Université de Lorraine, Institut Jean Lamour, CNRS, 54000 Nancy, France
[2] ITMO University, Saint Petersburg, Russia
[3] The Rockefeller University, New York, NY 10065, USA
[4] Department of Physics, Chemistry and Biology (IFM), Linköping University, Linköping 58183, Sweden
[5] ETH Zurich, Department of Physics, Institute for Quantum Electronics, Optical Nanomaterial Group, 8093 Zurich, Switzerland
[6] Ernst Ruska-Centre for Microscopy and Spectroscopy with Electrons, Forschungszentrum Jülich GmbH, 52425 Jülich, Germany
[7] Laboratoire de Chimie et Physique, Approche Multi-Echelle des Milieux Complexes, Univ. Lorraine, Metz, France
[8] Laboratoire de Physique des Gaz et des Plasmas, UMR8578 CNRS, Université Paris Sud, Université Paris Saclay, France
[9] Department of Gaseous Electronics, Jožef Stefan Institute, Jamova cesta 39, SI-1000 Ljubljana, Slovenia
[10] Max-Planck-Institut für Eisenforschung GmbH, Max-Planck-Straße 1, 40237 Düsseldorf, Germany
[11] Department of Physics, Aristotle University of Thessaloniki, GR-54124 Thessaloniki, Greece.
[12] Center for Interdisciplinary Research and Innovation, Aristotle University of Thessaloniki, Thessaloniki, Greece.
[13] Department of Physics, Engineering Physics and Astronomy, Queen's University, Kingston, Ontario K7L 3N6, Canada
[14] LORIA, Université de Lorraine, CNRS, INRIA, Nancy, France
[†] Corresponding authors : valentin.milichko@univ-lorraine.fr, alexandre.nomine@univ-lorraine.fr, uros.cvelbar@ijs.si

**Supplementary note 1: Characterization of bulk Cantor HEA target**

The HEA sample was analysed in transmission electron microscope operating in scanning mode (STEM, ARM200CF, JEOL Inc.). For EDS analyses of chemical composition, the sample was mounted in low-background Be holder. TEM sample was prepared by site-specific focus ion beam technique (FIB, Helios NanoLab 650i, FEI Inc.).

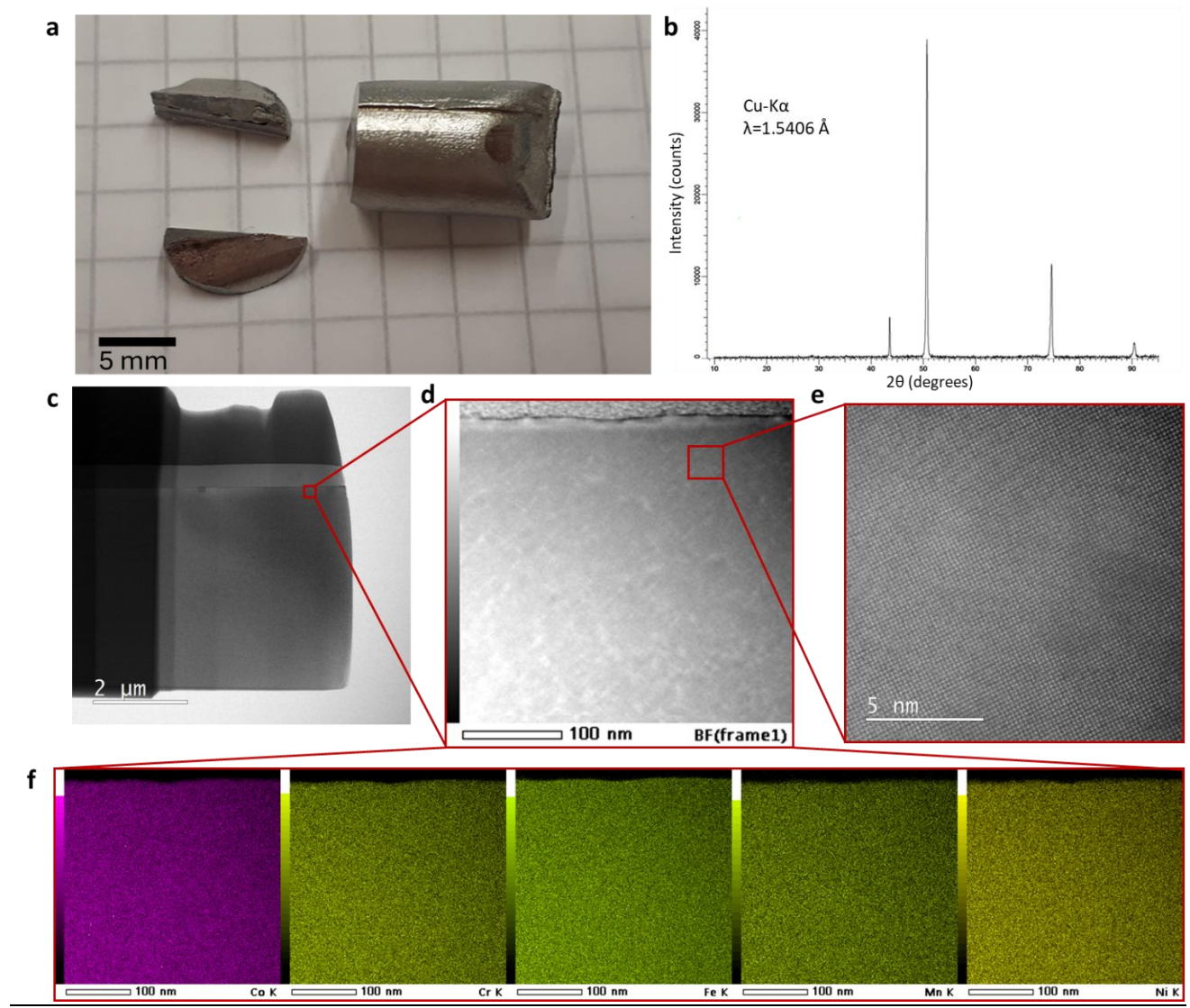

**Supplementary Figure S1.** Cantor Alloy, (a) photograph of ingot. b) XRD pattern showing diffraction peaks consistent with single-phase fcc crystal structure. (c-f) STEM analysis of the FIB-made lamella. (c) BF-STEM micrograph of HEA target, in-situ covered by protective layer to prevent surface oxidation. (d) HAADF-STEM micrograph of the HEA sample, with € magnified near-surface region. (f) STEM-EDS spatial distribution maps of principal components Co, Cr, Fe, Mn, Ni, showing homogenous distribution within the sample.

## Supplementary note 2: Optical measurements on HEA targets

**Ellipsometry:** The ellipsometric spectra are measured at two angles of incidence (50° and 70°) in 270-2100nm spectral range by using a phase modulated ellipsometer (UVISEL, HORIBA). Two ellipsometric angles $\psi$ and $\Delta$ were recorded at each wavelength. These angles are directly related to the ratio $\rho$ between the $r_p$ and $r_s$ Fresnel reflection coefficients:

$$\rho = \frac{r_p}{r_s} = \tan(\psi)\, e^{j\Delta}.$$

The ellipsometric spectra are shown in Supplementary Figure S2a,b. The material is sufficiently thick and opaque to be modeled as a semi-infinite medium. In this case, the complex dielectric function of the materials can be deduced from the following equation:

$$\varepsilon = \sin^2(\theta)\left(1 + \left(\frac{1-\rho}{1+\rho}\right)^2 \tan^2(\theta)\right),$$

where $\theta$ denoted the angle of incidence. The real part of the dielectric function (Supplementary Figure S2c) is negative revealing the metallic character of the material. The signal to noise ratio is degraded in the infrared part due to the spectral variation of the sensitivity of the ellipsometric sensor.

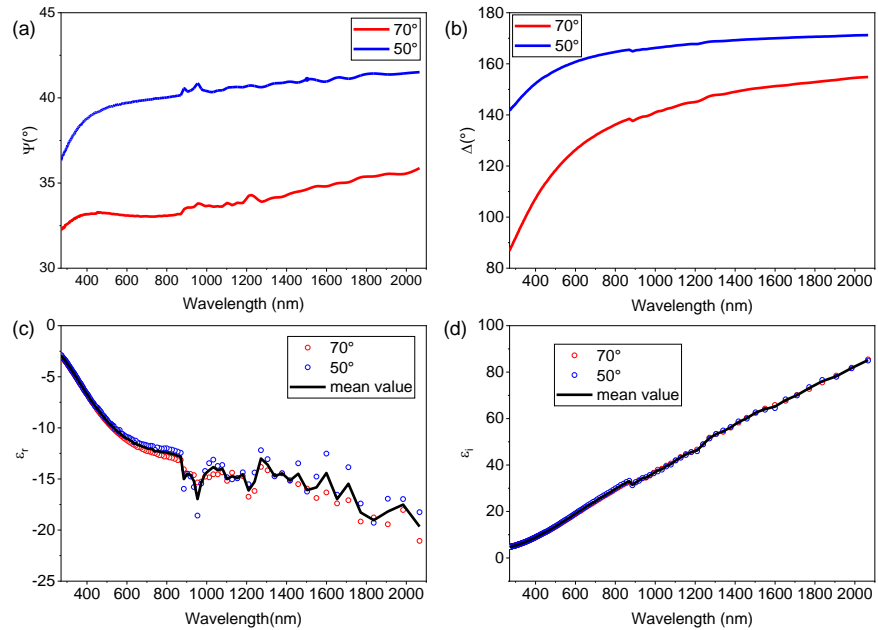

**Supplementary Figure S2.** (a) $\Psi$ and $\Delta$ ellipsometric spectra measured at angles of incidence of 50° and 70°. (c) Real part and (d) imaginary part of the dielectric function of the material measured at angles of incidence of 50° and 70°. The mean values are also reported (solid black curve).

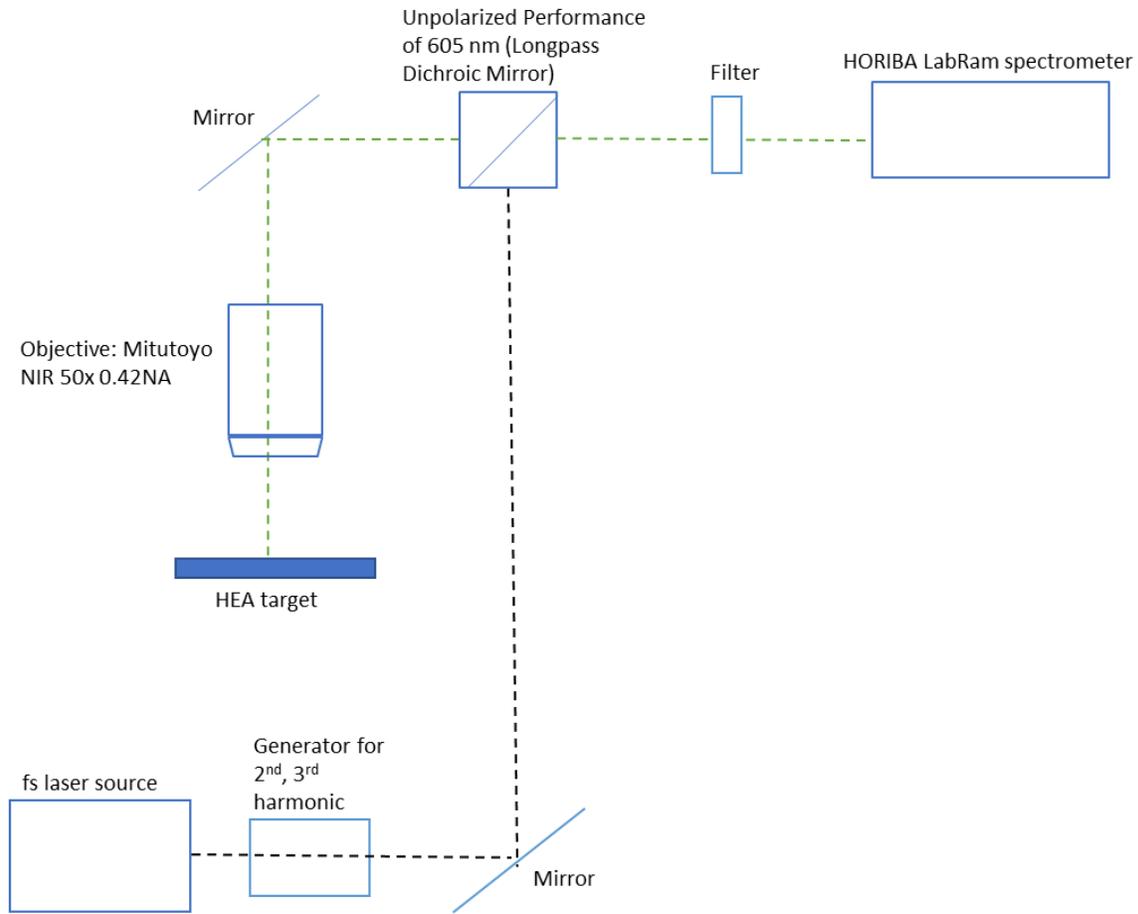

**Supplementary Figure S3.** Schematic illustration of optical set-up for analysis of SHG and broadband light emission in the reflection geometry. Femtosecond laser source: commercial femtosecond laser system (TEMA, Avesta Project), emitting 150 fs laser pulses at a wavelength of 1050, 525, and 350 nm with a repetition rate of 80 MHz. Mirrors: Thorlabs protected silver mirrors with 97.5% reflectance within the (0.45;2) µm region. Objective: Mitutoyo M Plan Apo NIR 50x (NA=0.42). Filter: Thorlabs FELH0600, FESH 0600 allowing to cut the pumping wavelength. Spectrometer: commercial confocal spectrometer Horiba LabRam with thermoelectrically cooled charge-couple device (Andor iDus InGaAs) and 150 g mm$^{-1}$ diffraction grating.

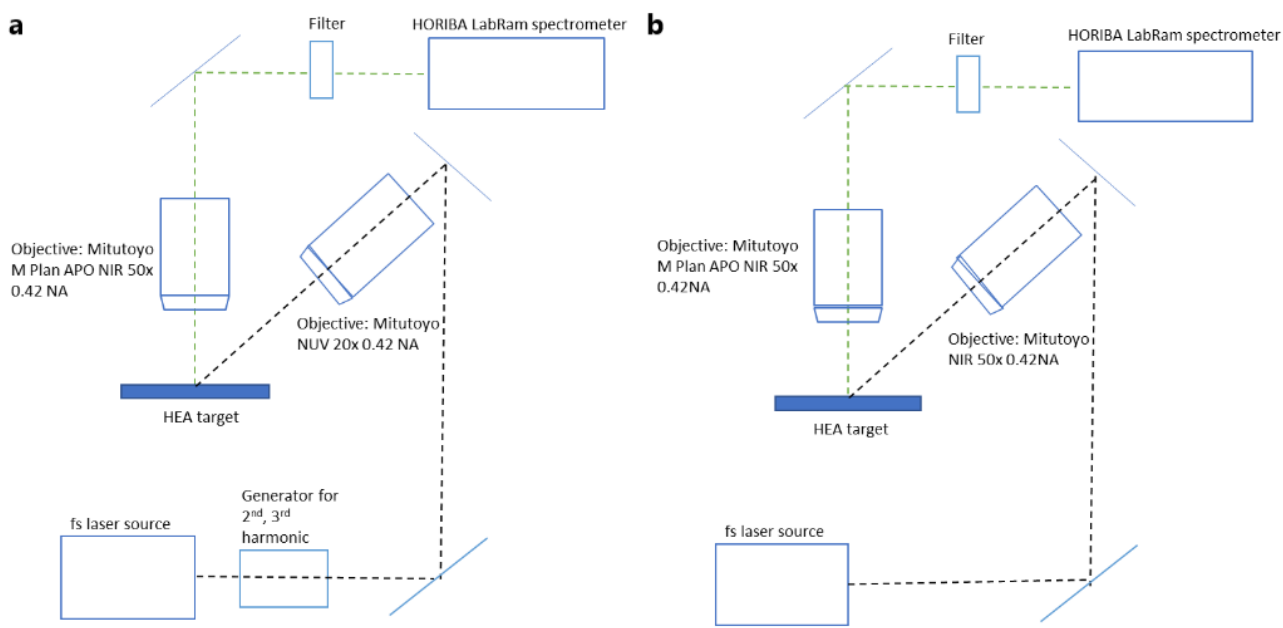

**Supplementary Figure S4.** Schematic illustration of optical set-up for analysis of a broadband light emission (a) and SHG (b) in the scattering geometry. Femtosecond laser source: commercial femtosecond laser system (TEMA, Avesta Project), emitting 150 fs laser pulses at a wavelength of 1050 (for SHG) and 525, 350 nm (to initiate the broadband light emission) with a repetition rate of 80 MHz. Mirrors: Thorlabs protected silver mirrors with 97.5% reflectance within the (0.45;2) µm region. Objectives: Mitutoyo M Plan Apo NIR 50x (NA=0.42) to collect the optical signal; Mitutoyo NUV 20x (NA=0.42), and Mitutoyo NIR 50x (NA=0.42) to initiate SHG and broadband light emission. Filter: Thorlabs FELH0600, FESH 0600 allowing to cut the pumping wavelength. Spectrometer: commercial confocal spectrometer Horiba LabRam with thermoelectrically cooled charge-couple device (Andor iDus InGaAs) and 600, 150 g mm$^{-1}$ diffraction grating.

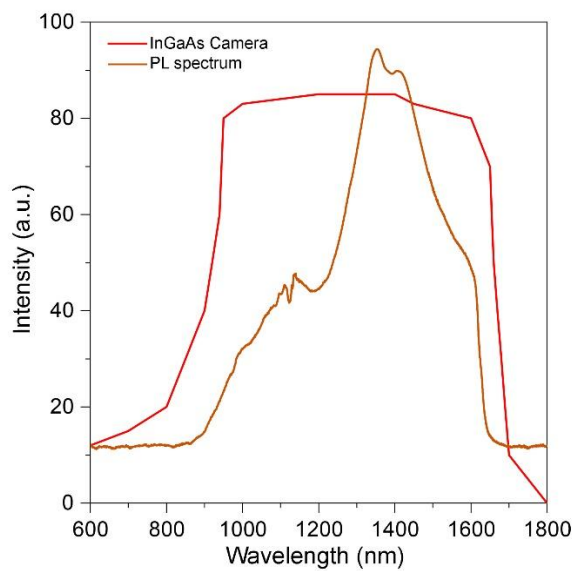

**Supplementary Figure S5.** Infrared detector (Andor iDus InGaAs) sensitivity compared to the broadband light emission from Cantor target.

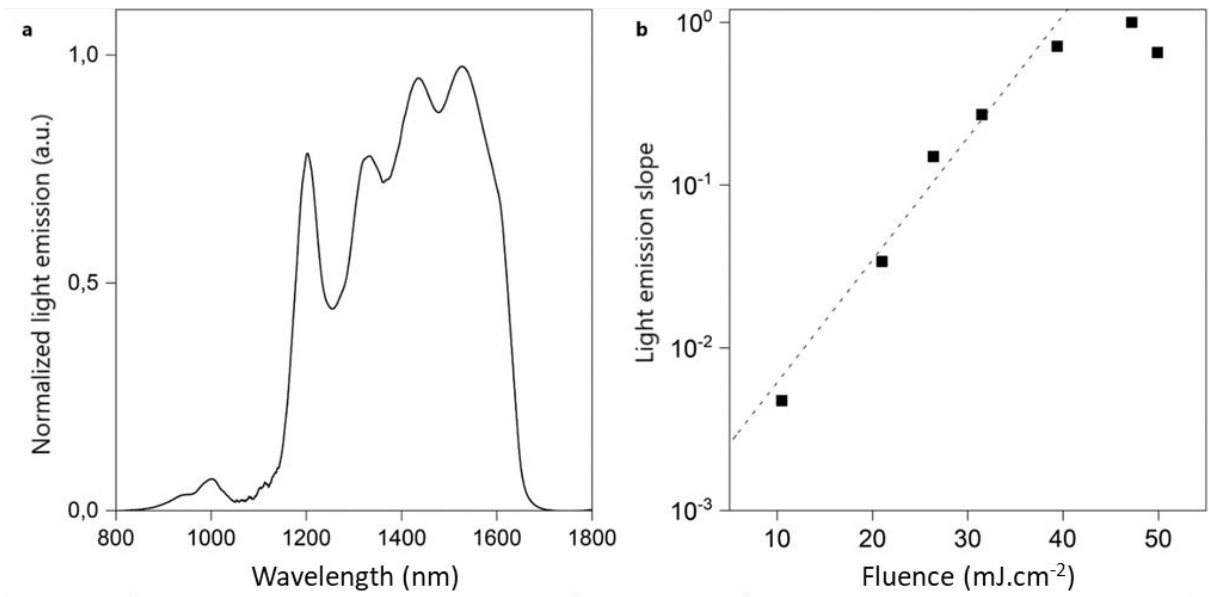

**Supplementary Figure S6.** (a) Normalized broadband light emission from the surface of AlCrFeNiTi$_{0.21}$ target, pumped by 525 nm light of a varied fluence (b), demonstrating app. 4 order of the slope.

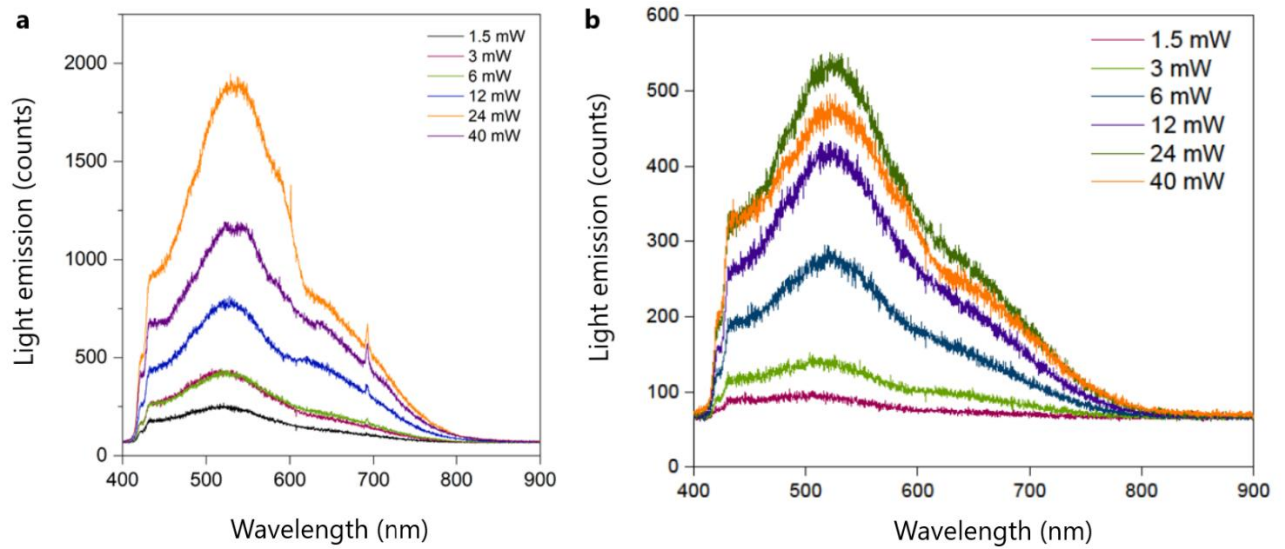

**Supplementary Figure S7.** (a) Broadband light emission from the surface of AlCrFeNiTi$_{0.21}$ target and (b) Cantor target, pumped by 350 nm light of a varied power (and analyzed under the same conditions).

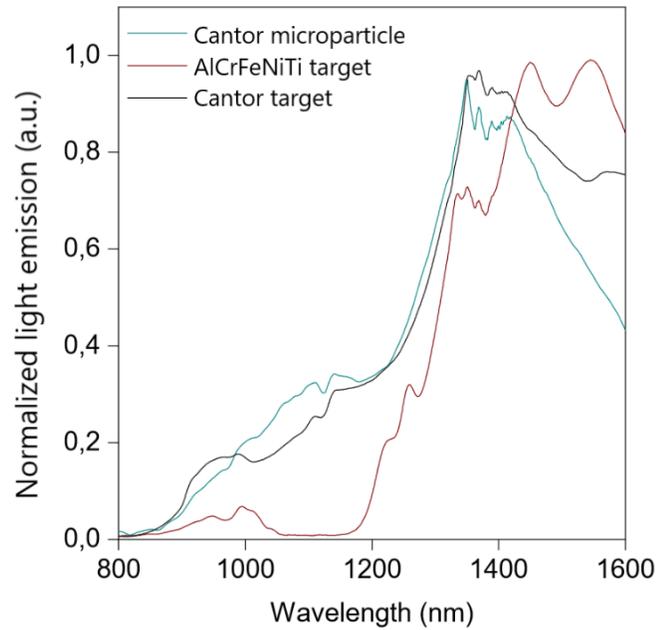

**Supplementary Figure S8.** Comparison of the normalized broadband light emission spectra from the AlCrFeNiTi$_{0.21}$, Cantor target, and Cantor alloy microparticle (from Supplementary note 4), pumped by 525 nm light.

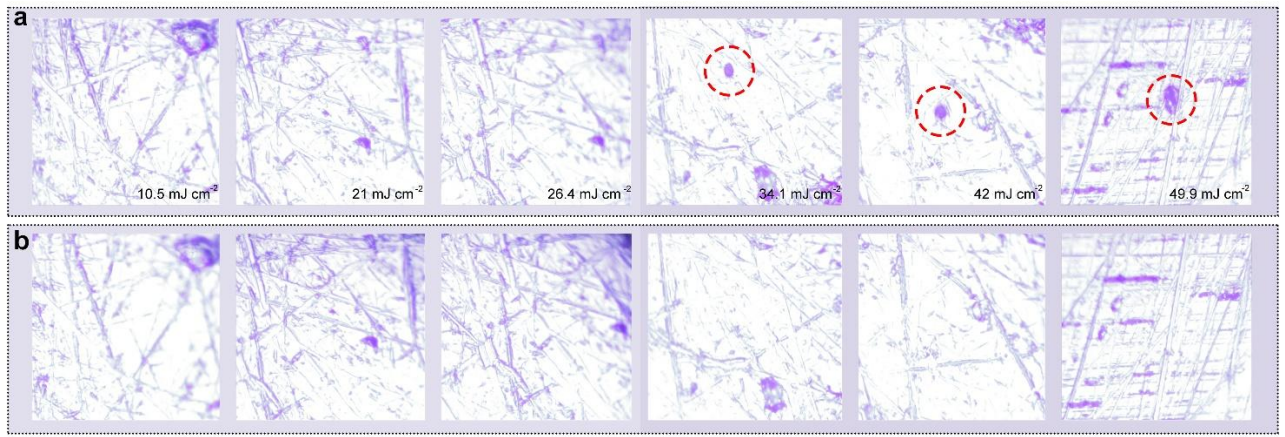

**Supplementary Figure S9.** Optical images allowing one to compare the surface of Cantor target before (b) and after (a) pumping it by 525 nm light of a varied fluence.

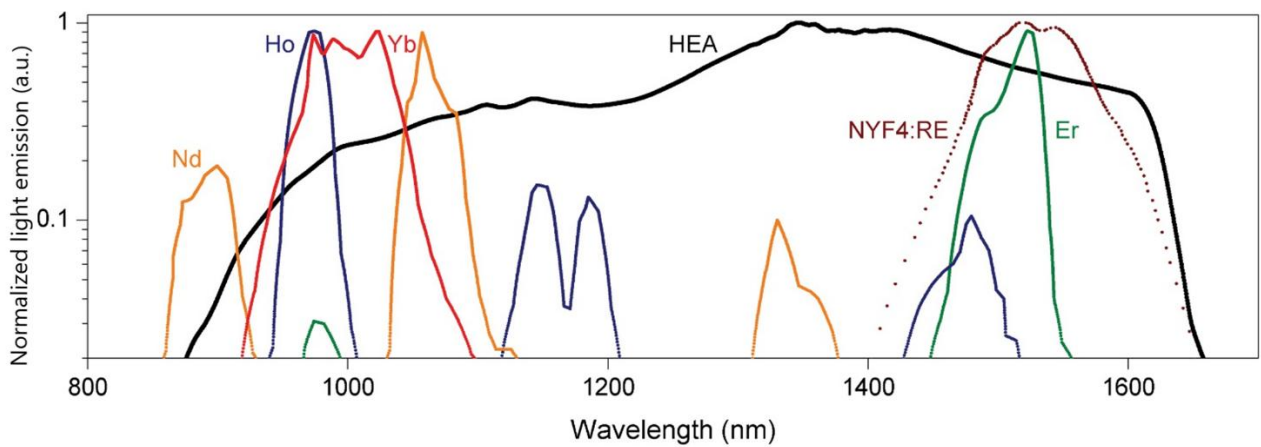

**Supplementary Figure S10.** Comparison of the normalized spectra of the broadband light emission from the Cantor target (pumped by 525 nm light) and light emission from rare-earth elements and lanthanides.[1]

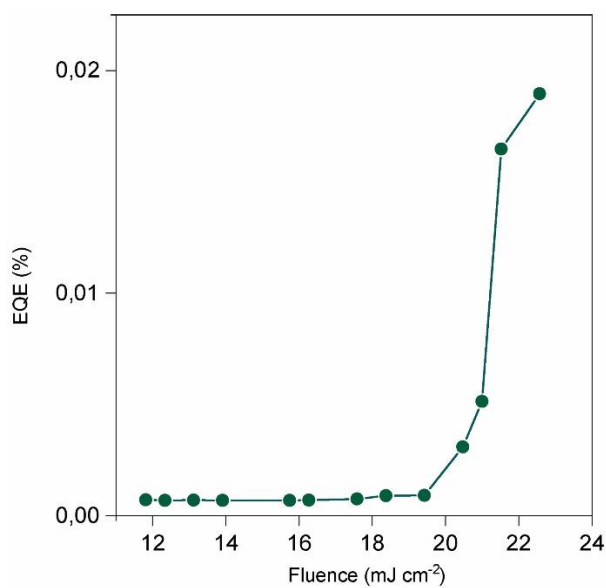

**Supplementary Figure S11.** Estimated quantum efficiency (in %) of the broadband light emission from the Cantor target (pumped by 525 nm light) over increased pumping fluence.

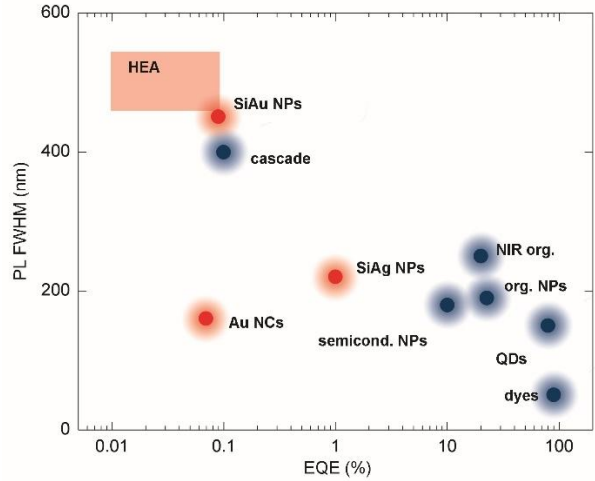

**Supplementary Figure S12.** External quantum efficiency related to full width at half maxima (FWHM) for different subwavelength emitters and Cantor: SiAu NPs[5], cascade nanostructure[6], Au nanoclusters[7], SiAg nanoparticles[8], semiconductor nanoparticles[9], near infrared organic emitters[10], organic nanoparticles[11], quantum dots[12], and dyes[13].

**Black-body emission:** The peak emission of HEA emission spectra is equal to 1380 nm 9Supplementary Figure S8). In the hypothesis of a black-body emission type, the temperature by the Wien law:

$$\lambda_{peak} = \frac{\sigma_w}{T}$$

with $\lambda_{peak}$, the peak emission wavelength, $\sigma_w$ the Wien's displacement constant (2.897$10^{-3}$ m·K), and T the Temperature. This lead to a temperature of 2100 K which overcomes the melting point of Cantor alloy ($T_m = 1607\ K$)[2,3] and AlCrFeNiTi ($T_m = 1718\ K$)[4] and makes already the hypothesis of an emission coming from blackbody emission rather unrealistic.

The blackbody emission radiation intensity as a function of the wavelength can be also calculated with the following formula:

$$I(\lambda) = \frac{2 \cdot h \cdot c^2}{\lambda^5} \cdot \frac{1}{e^{\frac{hc}{\lambda k_B T}} - 1}$$

with $\lambda$ the wavelength, $c$ the speed of light, $h$ and $k_B$ the Planck and Boltzmann constants respectively. The theoretical blackbody emission radiation spectra calculated at $T = 2100\ K$ and $T = T_m = 1607\ K$ are represented and compared with Cantor alloy emission spectra in Supplementary Figure S13. It is obvious that besides the inconsistency between the temperature corresponding the peak intensity and the melting temperature, the spectra shape of the Cantor alloy emission does not correspond to the one of a blackbody emission. Thereby, the hypothesis of an emission due to thermal effect can be ruled out.

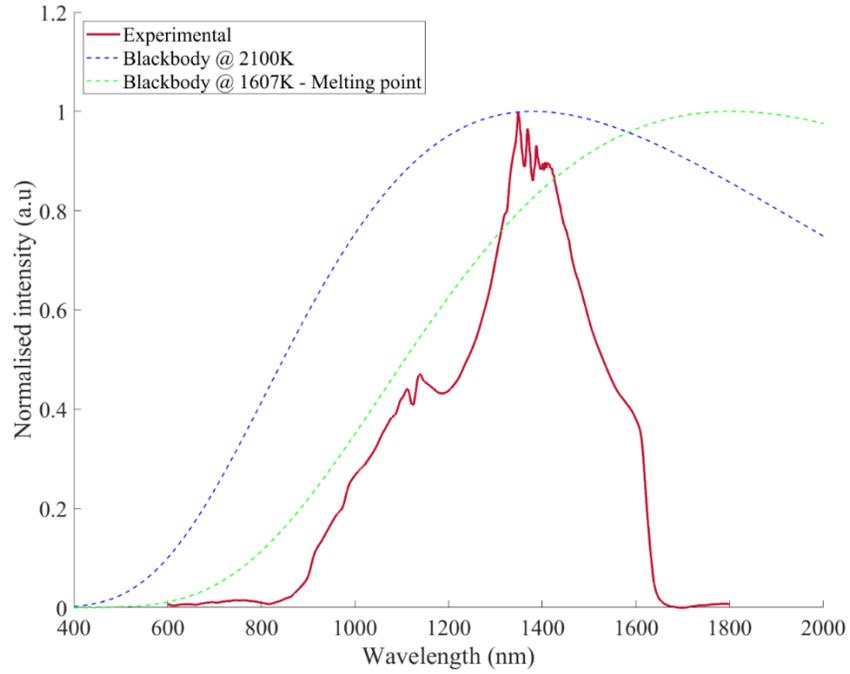

**Supplementary Figure S13.** Comparison of the experimental spectrum (red curve) of the broadband light emission from the Cantor target (pumped by 525 nm light) with theoretical black body emission spectra at 2100 K (blue dotted curve), corresponding to a maximum emission of 1400 nm, and at 1607 K (green dotted curve) corresponding to the melting point of Cantor alloy.

**Supplementary Note 3: Cantor microparticles**

**Synthesis:** The experimental setup consists in imersing Cantor alloy electrodes in liquid nitrogen and applying 10 kV and 500 ns pulses during 600 ns. DC voltage power supply (Technix SR15-R-1200–15 kV–80 mA) was connected to a solid-state switch (Behlke HTS-301-03-GSM) connected to one pin-electrode, while the other pin-electrode was grounded. The voltage rise time was 20 ns. The pulse frequency was set to 10 Hz during 30 minutes. The inter-electrode gap distance remained constant at 100 µm. Further details on the synthesis can be found in Hamdan *et al.*[14].

This process has the advantage to be easy to set up although it makes possible to reach extremely high cooling rates ($10^6 – 10^{12}$ K s$^{-1}$). When carried out in liquid nitrogen, it is possible to limit the oxidation and to synthesize HEA nanoparticles with a size varying from 0.1 to 2 µm. Compared to other synthesis techniques such as carbothermal shock[15], Electrical Discharge in Dielectric Liquid (EDDL) makes it possible to grow larger nanoparticles with a size varying in the range of 1/10 λ to 4 λ (λ being the incident wavelength equal to 1050 nm for SHG study and 520 nm for PL measurement) which is the most favorable range to observe non-linear responses enhanced by size.

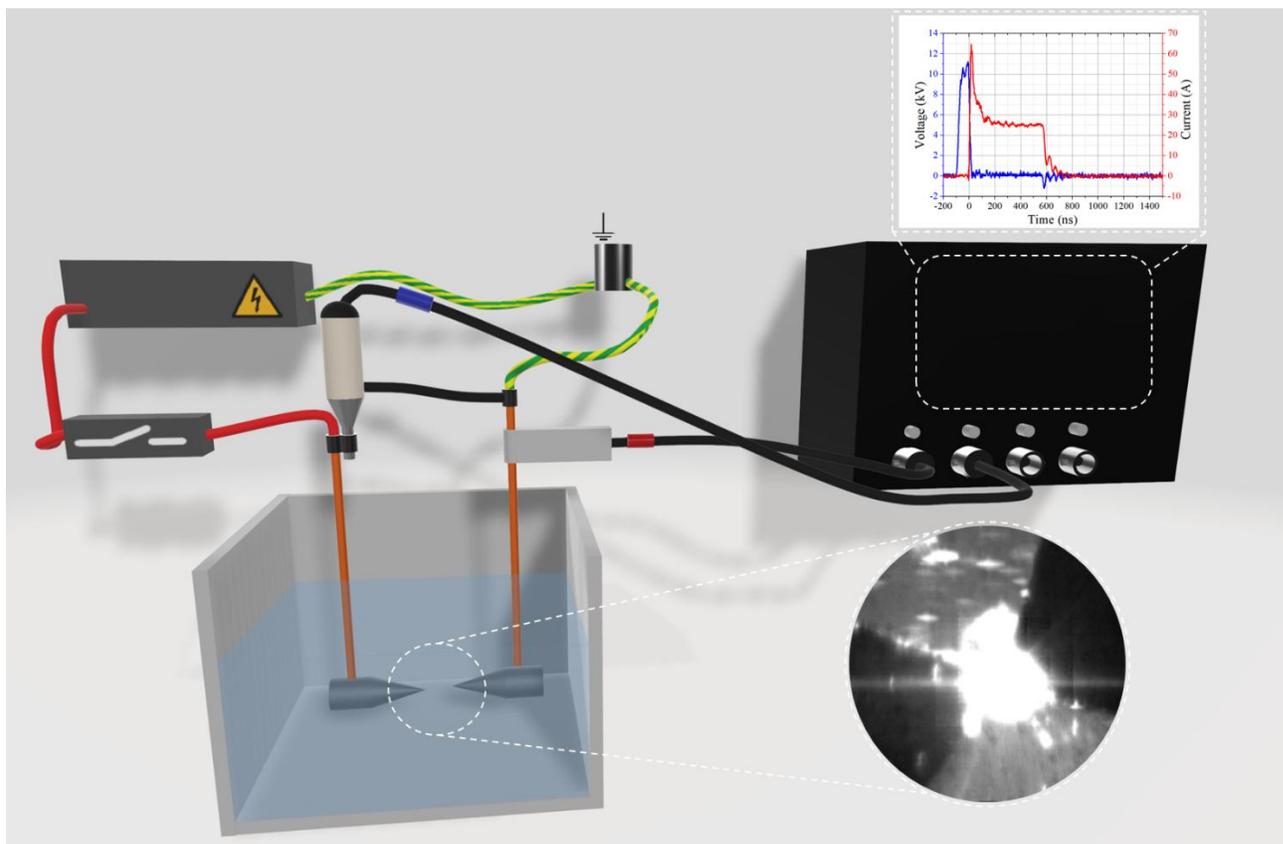

**Supplementary Figure S14.** Experimental setup for HEA nanoparticle fabrication based on Electric Discharges in Dielectric Liquid.

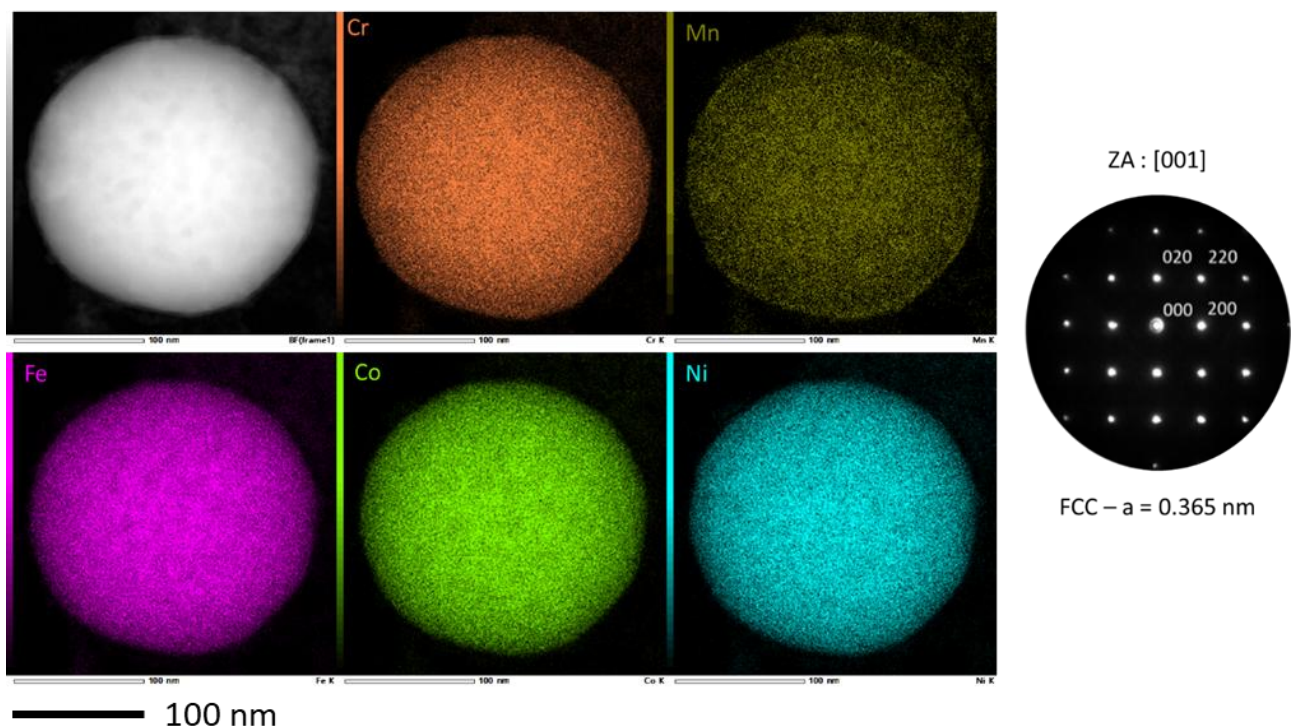

**Supplementary Figure S15.** HAADF-STEM micrograph of HEA nanoparticle, and corresponding EDS-STEM spatial distribution maps of principal elements Cr, Mn, Fe, Co, Ni. The selected-area electron diffraction (SAED), indexed for *fcc* structure, confirms single-crystal nature of the particle.

## Supplementary Note 4: Optical measurements on Cantor microparticles

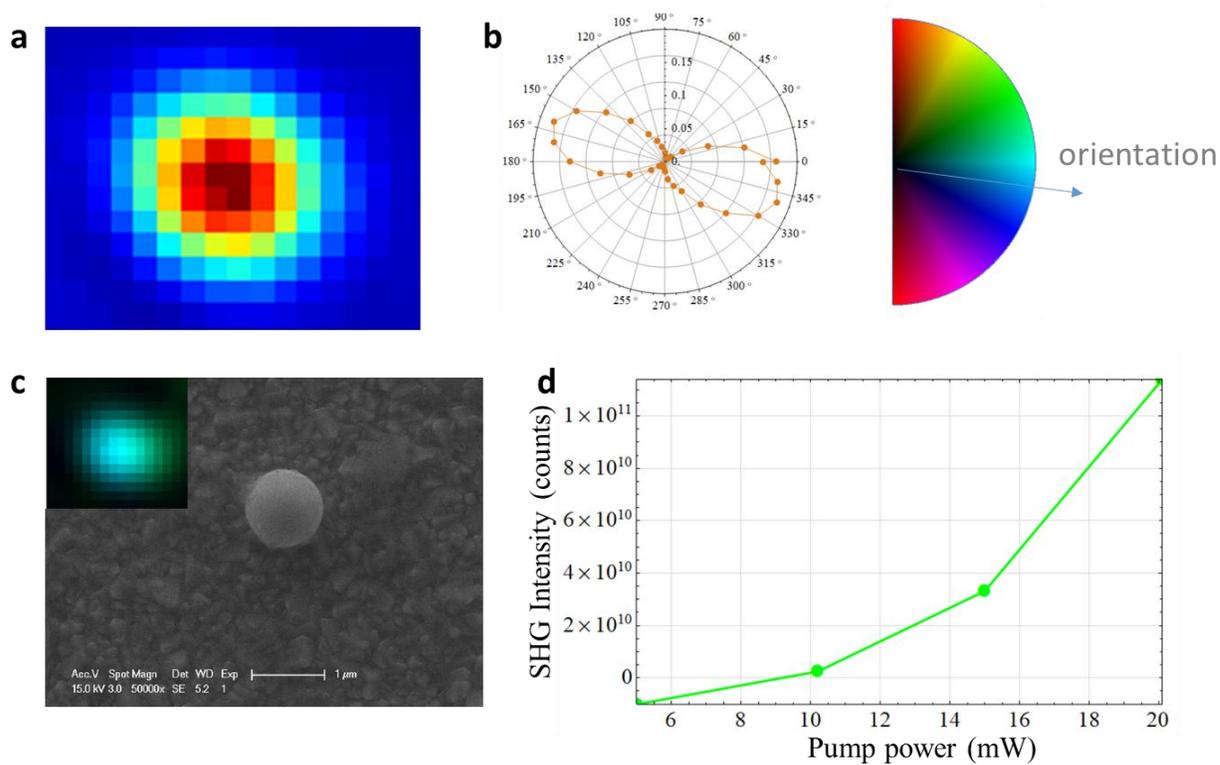

**Supplementary Figure S16.** (a) Imaging SHG analysis for single Cantor microparticle, pumped by 1000 nm light, revealing the corresponding polarization behavior (b) and quadratic slope (d) over pumping power. (c) SEM micrograph and optical image (inset) of the corresponding Cantor microparticle.

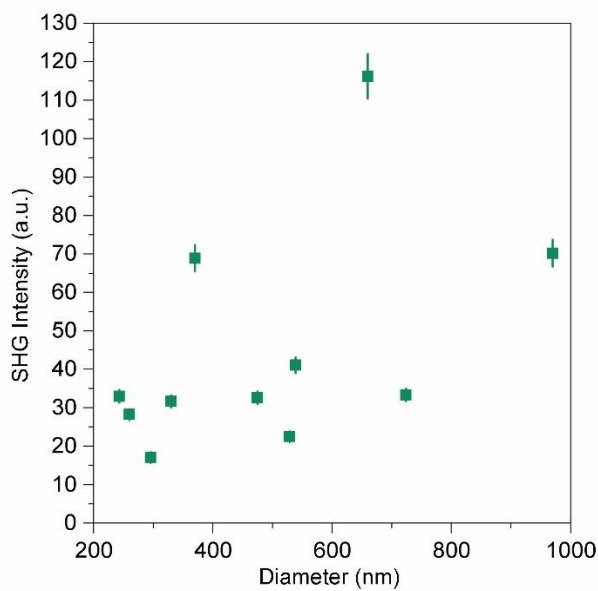

**Supplementary Figure S17.** Dependence of SHG intensity for single Cantor microparticles on their size, measured at the same experimental conditions.

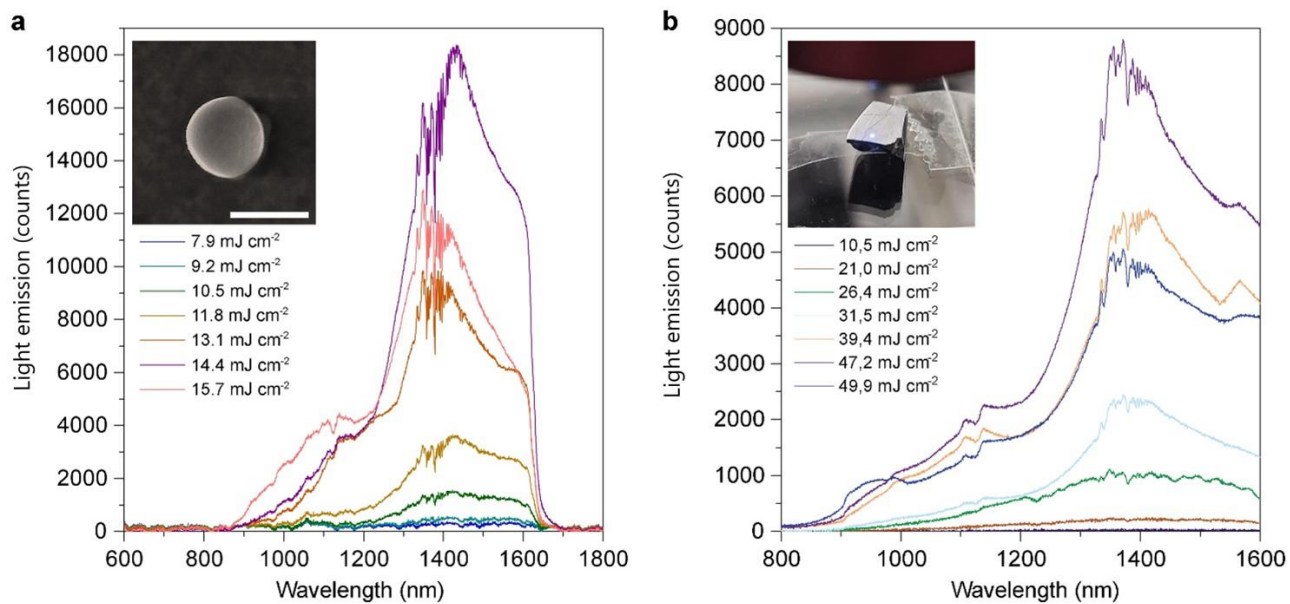

**Supplementary Figure S18.** Comparison of the spectra of the broadband light emission from the Cantor alloy microparticle (a), and the Cantor target (b), pumped by 525 nm light with different fluence. Inset: (a) SEM micrograph of the Cantor alloy microparticle, scale bar, 1 µm (a); (b) Optical image of the Cantor target piece.

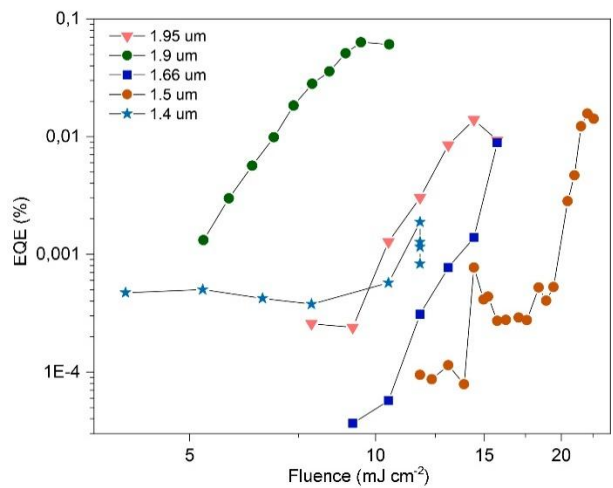

**Supplementary Figure S19.** External quantum efficiency of the broadband light emission from single Cantor microparticles with different diameters vs pumping fluence (525 nm), compared to that of Cantor target (Supplementary Figure S11).

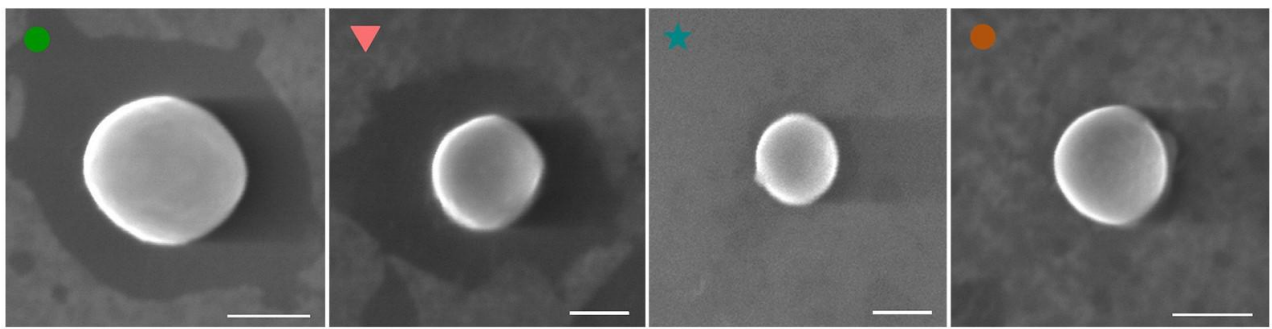

**Supplementary Figure S20.** Corresponding SEM micrographs of single Cantor microparticles (for Supplementary Figure S19). Scale bars, 1 µm.

**Supplementary note 5: DFT Modelling of electron energy levels**

The attained CLS for all the 80 atoms in the multi-component alloy supercell are presented in Supplementary Figure S21 as histograms. The room temperature stable elemental phases of respective element were used as reference, and CLS of the multi-component alloy was calculated, noting that for Mn we used an antiferromagnetic fcc structure as the reference. All atoms show a shift (CLS values different from 0), consistent with the changed electronic structure, there is for all elements a spread in observed CLS, which is the result of the varying chemical surroundings. The spread of CLS values for each element can be expected to correlate to an experimentally observed alloy broadening.

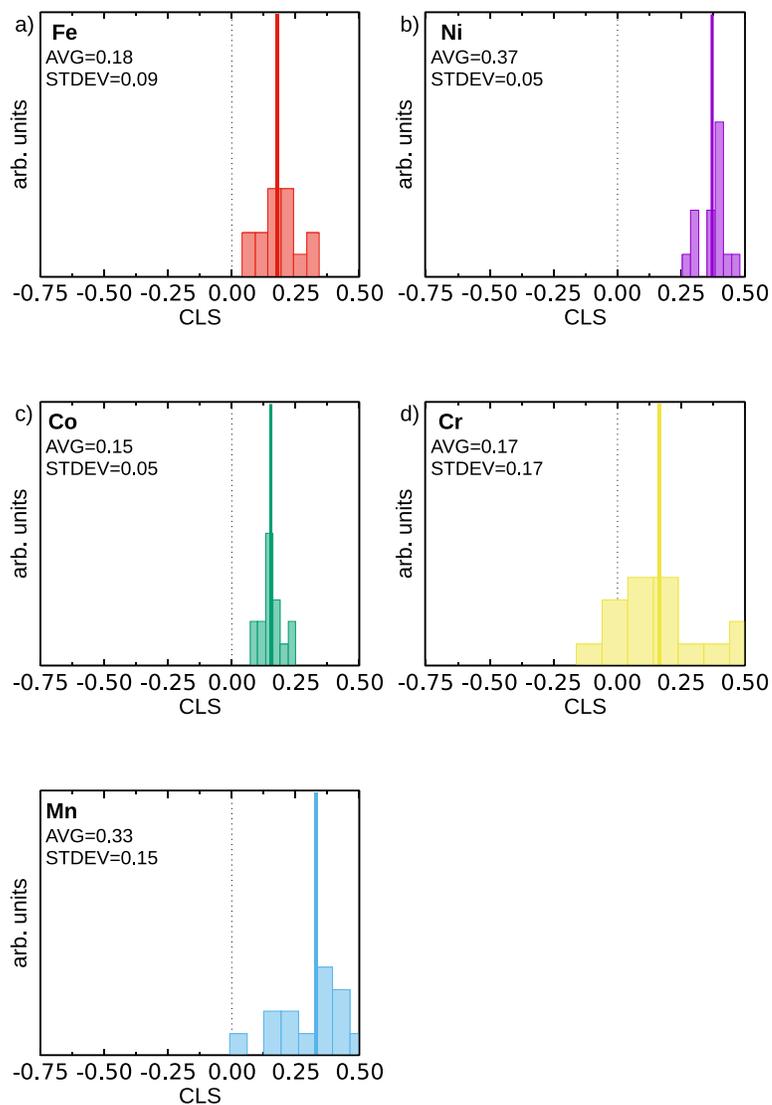

**Supplementary Figure S21.** Histograms of simulated XPS CLS for the 80 atoms of the DFT supercell, divided according to element. Average CLS indicated in a solid line.

**Supplementary note 6: DFT Modelling of lattice distortion**

For both $Al_{0.2375}Cr_{0.2375}Fe_{0.2375}Ni_{0.2375}Ti_{0.05}$ and $Fe_{0.2}Co_{0.2}Cr_{0.2}Mn_{0.2}Ni_{0.2}$ six random structures are considered, and the total energies of the final relaxed atomistic models are in a range of $\Delta E=$ 13 meV/atom by the use of GGA-PBE and 42 meV/atom with r2SCAN. It should be noticed that the energetically unfavorable atomistic configuration by GGA-PBE is found as the energetically favorable by r2SCAN, which means that all the model structures considered are energetically possible.

The displacement vectors of the relaxed atomistic configurations are calculated considering the final relaxed configuration with respect to the initial ones. By the use of OVITO, the change of the of the simulation supercell geometry is considered and performed a remapping of the final atomic positions into the reference simulation cell before calculating the displacement vectors.

In Supplementary Figure S22a the histogram of the displacement magnitudes of the six bcc $Al_{0.2375}Cr_{0.2375}Fe_{0.2375}Ni_{0.2375}Ti_{0.05}$ atomistic models are presented. It's clear that all the model structures are in agreement, since the vast majority of displacement amplitudes are less than 0.3 Å and their distributions are almost identical. The same conclusion is also validated in Supplementary Figure S23a, where the histogram of the displacement magnitudes of the $Al_{0.2375}Cr_{0.2375}Fe_{0.2375}Ni_{0.2375}Ti_{0.05}$ atomistic models relaxed by r2SCAN are presented. The model structure 5 is missing from the graph because it is not converged.

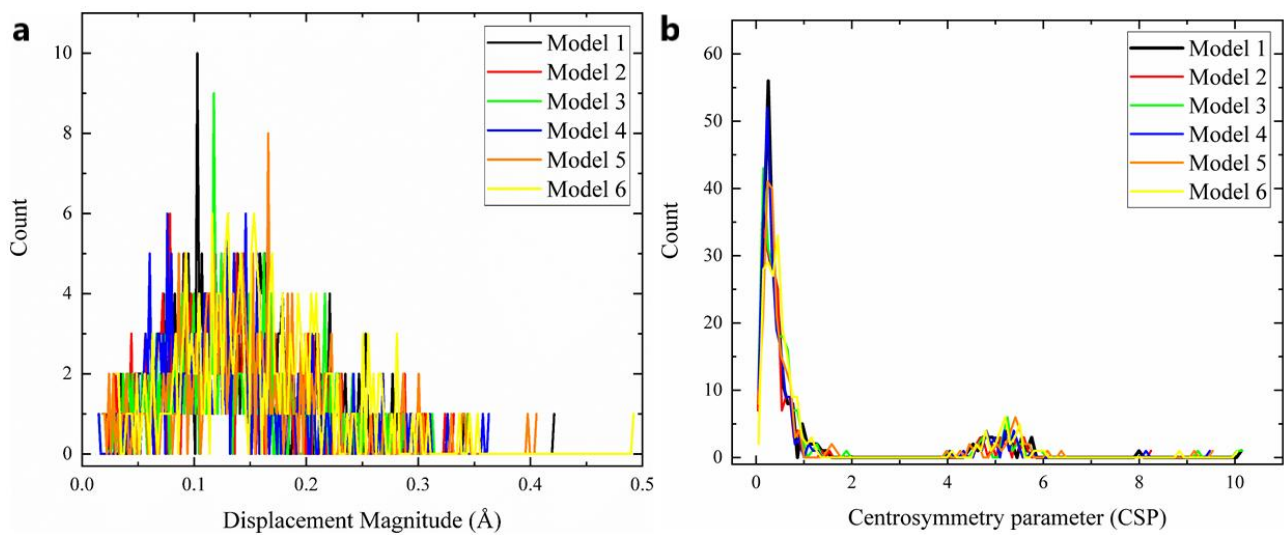

**Supplementary Figure S22.** The histograms of (a) the displacement magnitudes and (b) of the centrosymmetry parameter (CSP) of the six $Al_{0.2375}Cr_{0.2375}Fe_{0.2375}Ni_{0.2375}Ti_{0.05}$ atomistic models by the use of GGA-PBE.

In Supplementary Figure S22b histogram of the centrosymmetry parameter (CSP)[16] for each atom is presented, which is a useful measure of the local lattice disorder around an atom and can be used to characterize the deviation of an atomic position from the perfect lattice. The CSP value $P_{CSP}$, of an atom having N nearest neighbors (N=8 and N=12 for bcc and fcc respectively) is given by:

$$P_{CSP} = \sum_{i=1}^{N/2} |r_i + r_{i+N/2}|^2$$

where $r_i$ and $r_{i+N/2}$ are two neighbor vectors from the central atom to a pair of opposite neighbor atoms. For lattice sites in an ideal centrosymmetric crystal, the contributions of all neighbor pairs in this formula will cancel, and the resulting CSP value will hence be zero. The six bcc $Al_{0.2375}Cr_{0.2375}Fe_{0.2375}Ni_{0.2375}Ti_{0.05}$ atomistic models that are considered are present the same behavior, i.e. the deviation of the relaxed atomistic configurations from the perfect lattice are almost identical and very small variations between the six model structures are observed. This is concluded from both Supplementary Figure S22b and S23b, where the GGA-PBE and the r2SCAN approaches are used.

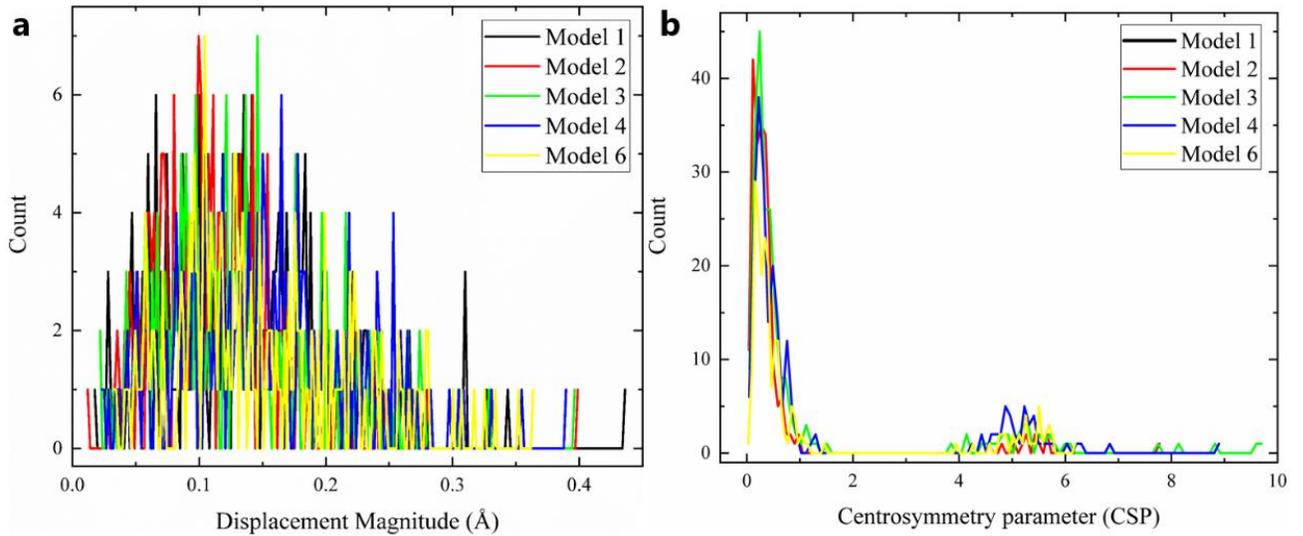

**Supplementary Figure S23.** The histograms of (a) the displacement magnitudes and (b) of the centrosymmetry parameter (CSP) of the six bcc $Al_{0.2375}Cr_{0.2375}Fe_{0.2375}Ni_{0.2375}Ti_{0.05}$ atomistic models by the use of r2SCAN. In both graphs the model structure 5 is missing because it is not converged.

In Supplementary Figure S24a the histogram of the displacement magnitudes of the six fcc $Fe_{0.2}Co_{0.2}Cr_{0.2}Mn_{0.2}Ni_{0.2}$ atomistic models are presented. It's clear that all the model structures are in agreement, since the vast majority of displacement amplitudes are less than 0.1 Å and their distributions are almost identical. It should be noted that displacement amplitudes of the fcc $Fe_{0.2}Co_{0.2}Cr_{0.2}Mn_{0.2}Ni_{0.2}$ model structures are smaller with respect the bcc $Al_{0.2375}Cr_{0.2375}Fe_{0.2375}Ni_{0.2375}Ti_{0.05}$ model structures. In Supplementary Figure S24b the histogram of the CSP for each atom of the six fcc $Fe_{0.2}Co_{0.2}Cr_{0.2}Mn_{0.2}Ni_{0.2}$ atomistic models is presented. The six fcc atomistic models that are considered present the same behavior, i.e. the deviation of the relaxed atomistic configurations from the perfect lattice are almost identical and very small variations between the six model structures are observed.

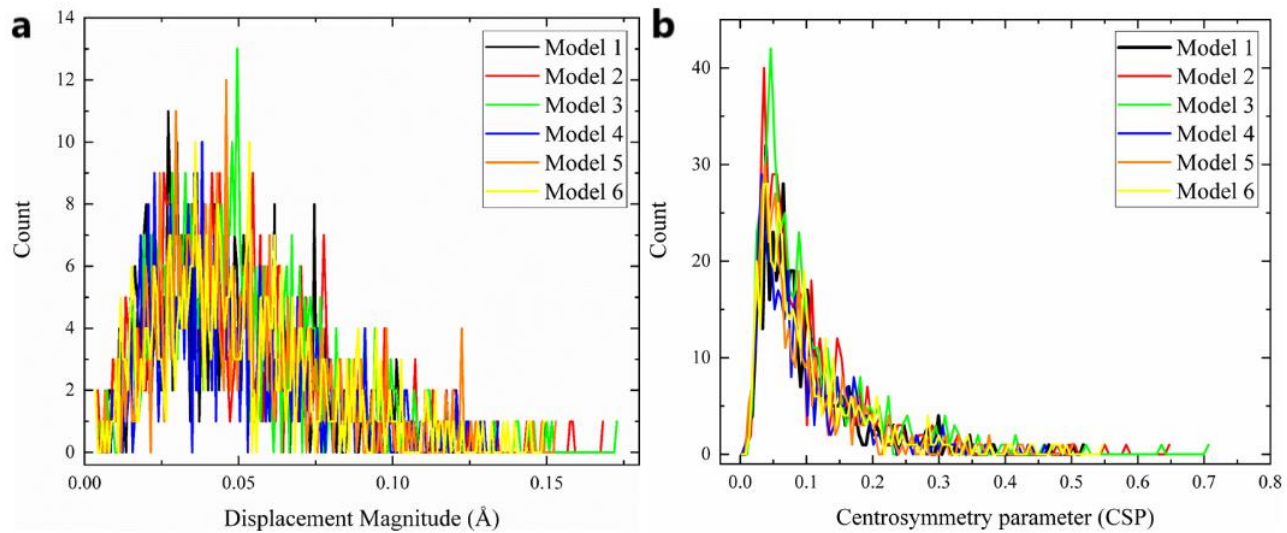

**Supplementary Figure S24.** The histograms of (a) the displacement magnitudes and (b) of the centrosymmetry parameter (CSP) of the six fcc $Fe_{0.2}Co_{0.2}Cr_{0.2}Mn_{0.2}Ni_{0.2}$ atomistic models by the use of GGA-PBE.

**Supplementary note 7: Sustainability Assessment**

The method to assess HEA sustainability is described in Nominé *et al.*[17]. The $CO_2$ footprint of the primary extraction of metal comes from Granta EduPack © Software, the companionality of the metals from Nassar *et al*[18], the Environmental-Social and Governance score from Lèbre *et al*[19], the production of metals from US Geological Survey[20].

The spectral range of emission of Silicon, InGaAs and $NYF_4$Yb:RE are estimated respectively from Meier *et al*[21], Hou *et al*[22] and Naczynski *et al.*[1]